\documentclass[amsmath,amssymb,prd,twocolumn]{revtex4}

\usepackage{graphicx}
\newcommand{\beq}{\begin{equation}}
\newcommand{\eeq}{\end{equation}}
\newcommand{\beqa}{\begin{eqnarray}}
\newcommand{\eeqa}{\end{eqnarray}}

\newcommand{\lexp}{\mathop{\langle}}
\newcommand{\rexp}{\mathop{\rangle}}

\def\d{\delta}

\def\dD{\delta_{\rm D}}

\font\BF=cmmib10

\def\k{{\hbox{\BF k}}}

\def\r{{\hbox{\BF r}}}

\def\v{{\hbox{\BF v}}}

\def\tk{\hat k}

 \def\la{\mathrel{\mathpalette\fun <}}
 \def\ga{\mathrel{\mathpalette\fun >}}
 \def\fun#1#2{\lower3.6pt\vbox{\baselineskip0pt\lineskip.9pt
        \ialign{$\mathsurround=0pt#1\hfill##\hfil$\crcr#2\crcr\sim\crcr}}}

\def\Mpc{\, h^{-1} \, {\rm Mpc}}

\def\kvecMpc{\, h \, {\rm Mpc}^{-1}}

\begin{document}

%
%
\title{Large-Scale Structure in Brane-Induced Gravity \\
I. Perturbation Theory}
%
%
\author{Rom\'an Scoccimarro}

\vskip 1pc

\affiliation{Center for Cosmology and Particle Physics, Department of
  Physics, \\  New York University, NY 10003, New York, USA} 

\begin{abstract}
%

We study the growth of subhorizon perturbations in brane-induced gravity using perturbation theory. We solve for the linear evolution of perturbations taking advantage of the symmetry under gauge transformations along the extra-dimension to decouple the bulk equations in the quasistatic approximation, which we argue may be a better approximation at large scales than thought before. We then study the nonlinearities in the bulk and brane equations,  concentrating on the workings of the Vainshtein mechanism by which the theory becomes general relativity (GR) at small scales. We show that at the level of the power spectrum, to a good approximation, the effect of nonlinearities in the modified gravity sector may be absorbed into a renormalization of the gravitational constant. Since the relation between the lensing potential and density perturbations is entirely unaffected by the extra physics in these theories, the modified gravity can be described in this approximation by a single function, an effective gravitational constant for nonrelativistic motion that depends on space and time. We develop a resummation scheme to calculate it, and provide predictions for the nonlinear power spectrum. At the level of the large-scale bispectrum, the leading order corrections are obtained by standard perturbation theory techniques, and show that the suppression of the brane-bending mode leads to characteristic signatures in the non-Gaussianity generated by gravity, generic to models that become GR at small scales through second-derivative interactions. We compare the predictions in this work to numerical simulations in a companion paper.

\end{abstract}



\maketitle

\section{Introduction}

The understanding of the current acceleration of cosmic expansion is one of the most pressing challenges in cosmology. While the cause of acceleration is presently not well understood, a substantial number of observations are being planned in the near future that will try to solve this puzzle. 

Cosmic acceleration can be explained in at least two different ways. Since in general relativity (GR) gravitational effects can be generated by pressure in addition to energy density (being, respectively, space-space and time-time components of the stress-energy tensor), the so-called {\em dark energy} solution to cosmic acceleration postulates a new contribution to the stress-energy tensor that dominates at late times and effectively has an equation of state where pressure is comparable to minus energy density (in natural units), leading to a repulsive force that fuels acceleration. The simplest, best motivated, such model in agreement with current observations is the cosmological constant.

Another possible explanation is that cosmic acceleration signals a breakdown of GR at present cosmological scales, rather than a new contribution to stress-energy. In this case, gravity is modified in the infrared (large-distance) regime to provide acceleration even though the universe is dominated by non-relativistic matter at late times.

Mapping the expansion history of the universe from deceleration to present acceleration typically cannot distinguish between these two possibilities, as e.g. the Friedmann equation in a theory of modified gravity can always be well approximated by a GR Friedmann equation with a dark energy component with suitable equation of state. The key is to add information on the growth of structure~\cite{2004PhRvD..69d4005L}, which results from a competition between gravitational attraction and the expansion history: different theories of gravity will result in different growth of structure for a fixed (observed) expansion history. By now, a large number of tests along these lines have been proposed in the literature~\cite{2001PhRvD..64h3004U,2004PhRvD..69d4005L,2004astro.ph..9224B,2005PhRvD..71h3004S,2005PhRvD..71b4026S,2006ApJ...648..797B,2006PhRvD..74b3512K,2006PhRvD..74d3513I,2007PhRvD..75b3519H,2007PhRvL..98l1301K,2007PhRvD..76b3507C,2007astro.ph..1782F,2008PhRvD..77j3513D,2008PhRvD..78b4015B,2008PhRvD..78b4015B,2008PhRvD..78j3509F,2008PhRvD..78f3503J,2008PhRvD..78d3514A,2008PhRvD..78d3002S,2008MNRAS.390..131A,2008arXiv0807.0810S,2008arXiv0809.3791Z,2008arXiv0812.0002S,2008arXiv0812.2244A,2009ApJ...690..923Z,2009JCAP...01..048S,2009arXiv0901.0919D}, and will be pursued in upcoming surveys.

Modifying gravity at large scales is a difficult task, and so far there has been limited success in finding consistent theories. GR is a very constrained theory, since it follows from requirements of locality in 3+1 dimensions, massless spin two graviton and general covariance~\cite{1955PhRv...98.1118K,1970GReGr...1....9D,1975AnPhy..89..193B,1986PhRvD..33.3613W,2003flog.book.....F}. Any modification to it entails new degrees of freedom. When allowing for complex models of dark energy, however, the meaning of what constitutes a modification of gravity can become nontrivial to define. For example, many theories usually considered in the cosmology literature as modification of gravity can be written as GR plus a (possibly non-minimally coupled) scalar field, i.e. some form of dark energy (a notable example in this class are $f(R)$ theories~\cite{2004PhRvD..70d3528C,2007PhRvD..76f4004H}). In such scalar-tensor theories, where only a scalar degree of freedom is added to GR, it is a matter of choice how to describe the physics (since both are equivalent). A more restrictive definition of gravity modification is  to define departures from GR as  such  theories in which tensor perturbations propagate differently than in GR even in the absence of any contributions to the stress energy of the background, i.e. gravitons behave differently than in GR even in flat space~\cite{2008arXiv0811.2197N}. In this case one considers as gravity modification the  addition of scalars only if they change the properties of the spin-2 sector, e.g. altering the speed of propagation of gravitons.

Finally, the most restrictive definition of modified gravity (and the one we adopt in this paper) involves modification in the number of degrees of freedom in the spin-2 sector as a requisite, e.g. extra graviton polarizations, or many gravitons. Examples include theories that realize the idea of massive gravity, e.g. large-scale extra-dimensions~\cite{2000PhLB..485..208D,2007JHEP...05..045K,2008PhRvL.100y1603D,2008JCAP...02..011D},  degravitation~\cite{2007PhRvD..76h4006D} and bigravity~\cite{2002PhRvD..66j4025D,2007PhRvD..76j4036B}. These theories generically have a scalar-tensor {\em regime}, where the predictions can be approximated by the addition of one or more effective scalars (e.g. the zero-helicity component of the graviton/s), but the theory in full is not equivalent to an arbitrary dark energy model. The best-known example is the DGP model~\cite{2000PhLB..485..208D}, in which the large distance behavior involves from the 4D point of view an infinite number of massive spin-2 states that describe the fact that the theory is fundamentally five-dimensional.  This class of theories also have more practical advantages. As already mentioned, general covariance imposes very powerful constraints on the nonlinear couplings of spin-2 fields. Requiring consistency leads to strongly constrained theories that are easiest to test (or rule out), since they contain only a small number of free parameters (instead of free functions). See e.g.~\cite{2007NuPhS.171...88G} for more discussion along these lines.

From the point of view of cosmology, the extra degrees of freedom in the metric that behave as scalars (in the sense of the 3+1-SVT decomposition, see below) are the ones that lead to readily observable consequences. To be a viable theory of gravity, however, any extra scalar degree of freedom must be suppressed in the solar system, where classical tests (light deflection in combination with non-relativistic motion) have shown GR to be a very good approximation. 

To date perhaps the most promising mechanism for such a suppression is a generalization of the Vainshtein effect found in massive gravity~\cite{1972PhLB...39..393V}, where  extra degrees of freedom become nonlinear through their self-interactions and decouple from matter, which then leads to the recovery of GR~\footnote{A somewhat similar effect, the so-called ``chameleon" mechanism~\cite{Mota:2003tc,2004PhRvL..93q1104K,2004PhRvD..69d4026K}, can happen in scalar field dark energy where the mass of the extra scalar depends on the local density, thus nonlinear effects due to coupling of the scalar to matter (instead of self-interactions) leads to a suppression of the scalar at small scales. Having a chameleon mechanism imposes strong constraints on the shape of the potential, i.e. it is not a generic feature of scalar field dark energy.}. The Vainshtein mechanism happens under rather generic conditions and plays a role in various theories of modified gravity~\cite{2002PhRvD..65d4026D,2006NJPh....8..326D,2008arXiv0811.2197N,2009arXiv0901.0393B}. Being a nonlinear effect, the scale at which it happens is related to the amplitude of the sources; for stars such as the sun the Vainshtein scale is tiny by  cosmological standards, but for cosmological sources it is expected to be in the interesting few Mpc range~\cite{2004PhRvD..69l4015L}. Therefore, at scales smaller than the Vainshtein scale these theories will be consistent with GR, and thus difficult to distinguish from dark energy by cosmological observations, i.e. the expansion history plus growth tests mentioned above would fail~\footnote{Below the Vainshtein scale of the sun (of order 100 parsecs) it may be possible still to test these theories by very precise measurements of anomalous precession of solar system bodies~\cite{2003PhRvD..67f4002L} and millimeter-size changes in the moon's orbit~\cite{2003PhRvD..68b4012D}. Remarkably, such tests are well within reach, see e.g.~\cite{2008PASP..120...20M}.}.

Such considerations show that probes of cosmic acceleration that are sensitive to small scale perturbations may yield no (or weak) tests on the cause of the expansion history, and could be mistakenly interpreted in favor of dark energy if the possibility of nonlinear effects in the modified gravity sector is ignored.   It is therefore important to look for signatures of such nonlinearities in modified gravity theories that can be used as a diagnostic in future observations. That's in part the motivation for this paper and its companion paper II~\cite{LSSbrane2}.

In this paper we study the growth of structure in brane-induced gravity in five dimensions~\cite{2000PhLB..485..208D}, known as the DGP model, which presents all the ingredients discussed above. This theory belongs to a broad  class of modified gravity theories due to large-scale extra-dimensions~\cite{2007JHEP...05..045K,2008PhRvL.100y1603D,2008JCAP...02..011D}, and presents similarities with other realizations of the idea of massive gravity such as degravitation~\cite{2007PhRvD..76h4006D} and bigravity~\cite{2002PhRvD..66j4025D,2007PhRvD..76j4036B}. For theories that implement massive gravity through something akin to the Higgs mechanism see e.g.~\cite{2004JHEP...05..074H,2004JHEP...10..076D}.

The DGP model has two branches of cosmological solutions, the so-called normal and self-accelerating branches. The latter, in the presence of non-relativistic matter, makes a transition from early time deceleration to acceleration~\cite{2001PhLB..502..199D,2002PhRvD..65d4023D} with de Sitter expansion in the long-time limit. In this paper, we concentrate on the self-acelerating branch as an example of a theory of modified gravity that leads to acceleration, even though several lines of work have shown that this branch has significant problems related to the appearance of a ghost mode at the linear level~\cite{2003JHEP...09..029L,2004JHEP...06..059N,2006JCAP...08..012D,2006PhRvD..73d4016G,2006JHEP...10..066C,2007PhRvD..76j4041I,2008arXiv0811.2197N}, and thus it may not be regarded as a viable description of our universe at all scales, particularly when quantum effects are included~\footnote{Strictly speaking, as far as I know, the linear theory ghost has been found in the de-Sitter limit, i.e. for $Hr_c=1$. In the realistic cosmological case where $H$ is time dependent, the decomposition into Kaluza-Klein modes often used does not hold as the background metric depends explicitly in time and the extra-dimension coordinate.}. It's worth pointing out, however, that a ghost is not a necessary ingredient to obtain self-accelerated solutions~\cite{2008arXiv0811.2197N}.

On the other hand, as we discuss below, such issues do not seem to impact the cosmological solutions we study here, in the sense that cosmological perturbations that respond to a standard fluctuation spectrum source do not show any pathological behavior. In addition, and most importantly, it is expected that the techniques developed here will be widely applicable to the study of cosmological perturbations in other (consistent) large-distance modifications of gravity. We thus take the DGP model in the spirit of the simplest example of theories (parametrized by a single free parameter, as $\Lambda$CDM dark energy) where observational signatures that can help determine the cause of cosmic acceleration in the near future can be addressed most easily. The DGP model itself is strongly constrained by current observations, see e.g.~\cite{2002PhRvD..66b4019D,2006PhRvD..74b3004M,2007PhRvD..75f4003S,2008PhRvD..78j3509F,2007PhRvD..76f3503W}.

Previous work in the literature discussed gravitational instability in this model in different approximations: \cite{2004PhRvD..69l4015L} obtained the quasistatic linear and nonlinear evolution in the spherical approximation, while \cite{2006JCAP...01..016K} developed the linear quasistatic approximation further by going beyond spherical symmetry and including an explicit treatment of the bulk perturbations, which justified the assumptions in~\cite{2004PhRvD..69l4015L}. To follow perturbations of wavelength larger than the Hubble radius one must go beyond the quasistatic approximation; this was first done in  \cite{2007PhRvD..75f4002S} using a scaling ansatz and later corroborated by~\cite{2008PhRvD..77h3512C} using an exact numerical solution of the linear equations. Except for~\cite{2004PhRvD..69l4015L} all these treatments work in linear perturbation theory using the master variable formalism developed in~\cite{2000PhRvD..62h4015M,2002PhRvD..66j3504D} to solve for the bulk perturbations. 

In this paper we extend previous work in several directions. First, we derive the linear solutions without relying on the master variable formalism, following the standard treatment of cosmological perturbations in GR after choosing suitable perturbation variables based on  the properties of gauge transformations along the extra dimension~\footnote{Our treatment here has some overlap with~\cite{2007PhRvD..75h4040K}, which also does not rely on the master variable formalism, and first wrote down the nonlinear equations for the brane-bending mode in the cosmological case. We go beyond their treatment by including the effects of normal derivatives and solving for the nonlinear equations and their implications for statistical quantities such as the power spectrum and bispectrum. For another recent approach to the nonlinear power spectrum see~\cite{2009arXiv0902.0618K}.}. Next we tackle the nonlinear case, focussing on the recovery of GR by the Vainshtein mechanism using a resummation technique to obtain the modified Poisson equation propagator. Finally, we apply these results to the calculation of the nonlinear power spectrum and the corrections induced in the bispectrum due to the nonlinear dynamics in the modified gravity sector. We then comment on the impact of these results on looking for modified gravity in observations. In a companion paper~\cite{LSSbrane2}, we develop an N-body code and test some of the analytic predictions developed here. For similar work on N-body simulations see~\cite{2009arXiv0905.0858S}.

\section{Five-Dimensional Metric Perturbations}

We start by reviewing metric perturbations in the context of braneworlds, in which the observable universe is a 3+1 brane embedded in a 4+1 dimensional spacetime. Most of this section contains well known results (see e.g.~\cite{2000PhRvD..62f4022K,2002hep.th....5220R} for reviews), but it serves to fix our notation, explain our unusual gauge choice, and motivate our strategy to solve the bulk equations described in section~\ref{LPbulk} below. 

\subsection{SVT Decomposition}

In this paper we are interested in the scalar perturbations to the metric as seen  by an observer in the brane. Following the standard SVT (Scalar, Vector, Tensor) decomposition (see e.g.~\cite{1984PThPS..78....1K,1992PhR...215..203M}) slightly generalized for one extra dimension, we can write in general for the scalar part of the metric,

\beqa
ds^2&=&\ \ \ {\rm e}^{2\beta} (1+2\phi)\, dt^2 +2 \, \nabla_iB \, dx^idt 
\nonumber \\ & & 
- \Big[{\rm e}^{2\alpha} (1-2\psi)\delta_{ij}+(\nabla^2_{ij}-\frac{\delta_{ij}}{3}\nabla^2)A \Big]\, dx^i dx^j 
\nonumber \\ & & 
-\, (1+\delta g_{44})\, dw^2 + 2\nabla_iC\, dx^i dw + 2F \, dt dw ,
\nonumber \\ & & 
\label{5Dmetric}
\eeqa

where $x^A=(t,x,y,z,w)$ with $A=0,\dots,4$,  $w$ is the coordinate on the extra (4th spatial) dimension with $w=0$ denoting the unperturbed brane position, and $\alpha(t,w)$ and $\beta(t,w)$ characterize the background solution

\beq
{\rm e}^{\alpha} \equiv  a+ \varepsilon\ |w|\, \dot{a}, \ \ \ \ \ 
{\rm e}^{\beta} \equiv 1+ \varepsilon\ |w|\, \frac{\ddot{a}}{\dot{a}},
\eeq

where $a(t)$ is the scale factor, and dots denote time derivatives. We assume $Z_2$ symmetry, all metric perturbations except $C,F$ are functions of $|w|$, $C,F$ instead are odd (henceforth we work with $w\ge0$). In the absence of perturbations the metric in Eq.~(\ref{5Dmetric}) reduces to the background solution found in~\cite{2001PhLB..502..199D}, with $\varepsilon=+1$ corresponding to the self-accelerated branch and $\varepsilon=-1$ to the normal branch. From now on we will concentrate on the self-accelerated branch, although occasionally we will discuss how some results change in the normal branch. Note that, although not obvious from Eq.~(\ref{5Dmetric}), the background bulk spacetime ($w>0$) is Minkowski, by making a change of coordinates this can be made explicit but the brane will have a complicated trajectory in that coordinate system, see e.g.~\cite{2002PhRvD..66j3504D}. Finally, we assume a spatially flat universe throughout this paper.

A few comments are in order regarding the form of Eq.~(\ref{5Dmetric}). The SVT decomposition implied there is based on the behavior of metric perturbations with respect to transformations of the 3 spatial dimensions of the brane. A bulk observer, could perform a similar SVT decomposition but with respect to 4 spatial dimensions, which leads to 4S, 6V and 5T modes that would correspond respectively to spin 0,1,2 in a local inertial frame. The 5T modes are the true physical degrees of freedom (only ones that propagate in the absence of sources) for a spin 2 massless graviton in 5D, the rest are induced by stress energy fluctuations. Out of the total of 15 modes, however,  2S and 3V modes can be set to zero by appropriate choice of gauge, leaving 10 physical modes (2S, 3V and 5T). However, because the background spacetime does not have symmetry under full 4D spatial rotations, these SVT modes do not evolve independently. Instead, it is easier to work with the 3D SVT decomposition modes which {\em do} evolve independently thanks to the fact that $w={\rm const.}$ hypersurfaces in the background space are homogeneous and isotropic. The price to pay is that some V and T modes in 4D project into S modes in 3D. That's why apart from the usual 4 scalar modes encoded in $\phi,\psi,A,B$, we have to deal with 3 new scalar modes described by $C,F,\delta g_{44}$. 

The 3D SVT decomposition divides metric perturbations into 7S (shown in Eq.~\ref{5Dmetric}), 6V and 2T modes. Gauge transformations can be used to set 3S and 2V modes to zero, leaving 4S, 4V and 2T physical modes (again, a total of 10 physical modes). For example, the scalar modes $B$ and $F$ in Eq.~(\ref{5Dmetric}) correspond to linear combination of 4D S and V modes (spin 0 and 1). The scalar mode $C$ corresponds to a linear combination of 4D S,V and T modes (spin 0,1,2). We now discuss briefly how to use suitable gauge transformations to eliminate 3 out of 7 scalar modes.

\subsection{Gauge Transformations and Gauge Choices}
\label{GTGC}

Gauge transformations for scalar perturbations can be written as
\beq
x^A \rightarrow x^A -\delta x^A,\ \ \ \ \ \delta x^A\equiv(\delta x^0, \nabla_i \delta x_s,\delta x^4).
\eeq
These three functions can be chosen to eliminate some of the scalar potentials in Eq.~(\ref{5Dmetric}), keeping in mind that there is an additional degree of freedom we have not yet discussed, corresponding to the brane position~\cite{2000PhRvL..84.2778G} which will not stay fixed at $w=0$ in the presence of stress energy fluctuations. To see which scalar metric perturbations can be set to zero we look at the transformation properties of metric perturbations under a scalar gauge transformation,

\beq
\phi \rightarrow \phi +\dot{ \delta x^0}+\dot{\beta}\, \delta x^0 +\beta'\, \delta x^4,
\label{phiGT}
\eeq
\beq
\psi \rightarrow \psi - \dot{\alpha}\, \delta x^0 - \alpha' \, \delta x^4,
\label{psiGT}
\eeq
\beq
A \rightarrow A+2\, {\rm e}^{2\alpha} \, \delta x_s
\label{AGT}
\eeq
\beq
B \rightarrow B + {\rm e}^{2\beta}  \, \delta x^0 -{\rm e}^{2\alpha}  \, \dot{\delta x_s}
\label{BGT}
\eeq
\beq
C \rightarrow C - \delta x^4 -{\rm e}^{2\alpha}  \, (\delta x_s)'
\label{CGT}
\eeq
\beq
F \rightarrow F + {\rm e}^{2\beta}  \, (\delta x^0)' - \dot{\delta x^4}
\label{FGT}
\eeq
\beq
\delta g_{44} \rightarrow \delta g_{44}+2\, (\delta x^4)'
\label{dg44GT}
\eeq
where dots and primes denote time and $w$ derivatives, respectively.

From these equations one can deduce a few useful gauges. The simplest to see, is a generalization of the longitudinal gauge, in this case one sets

\beq
A=B=C=0.
\label{Lgauge}
\eeq
This can always be obtained, since these conditions uniquely fix $\delta x_s,\delta x^0, \delta x^4$, respectively. An important property of this gauge is that {\em the brane is bent}, i.e. the position of the brane is not at $w=0$ in the presence of fluctuations, but rather at some function $w=w_b(x^\mu)$. 

The gauge that we use in this paper can be easily obtained from the longitudinal gauge, Eq.~(\ref{Lgauge}). We make a gauge transformation $w \rightarrow w - \delta x^4$ that only involves the fourth spatial coordinate (thus keeping $A=B=0$), such that $(\delta x^4)'$ is fixed by the condition $\delta g_{44} =0$ and the $w$-independent part of $\delta x^4$ is fixed by the condition that the brane is straight in this gauge, i.e. $w_b=0$. This can always be done because it is possible to set to zero the brane position with a $w$-independent shift of coordinates. Therefore, our gauge reads

\beq
A=B=\delta g_{44} =w_b=0
\label{gauge}
\eeq

Note that in going from the longitudinal gauge to our gauge we generate a nonzero $C$ which is given by,
\beq
C = \frac{1}{2} \int_0^w \delta \hat{g}_{44}(x^\mu,|w|)\, dw - |\hat{w}_b(x^\mu)| \, |w|',
\label{Cgen}
\eeq
where quantities with hats are in the longitudinal gauge, and the form of the second term respects that $C$ must me odd in $w$ while bringing back the brane to the unperturbed position. Close to the brane, say $w=0^+$, the first term in Eq.~(\ref{Cgen}) does not contribute and $C$ is directly given by the brane bending $ \hat{w}_b(x^\mu)$. Thus we may call $C$ the {\em brane-bending mode}. In geometric terms, $\nabla^2C$ determines the brane extrinsic curvature at subhorizon scales.

Since our gauge can be imposed at any $w$, all derivatives of $A,B,\delta g_{44}$ with respect to $w$ will vanish as well. In the following we present the equations of motion always in this gauge. Note also that the induced metric at the brane is given in terms of the five-dimensional metric by $g_{\mu\nu}(t,x^i)=[g^{(5)}_{\mu\nu}]_{w=0}$.

The most popular choice in the literature is the so-called Gaussian Normal (GN) gauge, where one fixes $\delta x^4$ as in Eq.~(\ref{gauge}), but instead fixes $(\delta x_s)'$ from the condition $C=0$ and $(\delta x^0)'$ from the condition $F=0$. To fully specify the gauge one fixes the $w$-independent part of  $\delta x_s$ by setting $A=0$ {\em at the brane} and that of $\delta x^0$ by setting $B=0$ {\em at the brane}. Since these two choices can only be taken at $w=0$, the $w$-derivatives of $A$ and $B$ in this gauge will not be zero at the brane (and elsewhere), and one must keep track of them. In this sense the GN gauge seems a bit cumbersome, although it is perfectly fine as a gauge choice~\footnote{The gauge choice in~\cite{2007PhRvD..75h4040K}, corresponding to $A=B=F=0$ in our notation, is not enough to completely fix the gauge, as only $\dot{\delta x^4}$ rather than $\delta x^4$ is fixed.}.

\subsection{ADM form}

To understand the role of the brane-bending more $C$ in geometric terms, see e.g.~\cite{2004JHEP...06..059N,2003JHEP...09..029L}, it is useful to cast the metric in ADM form~\cite{2008GReGr..40.1997A}, 

\beq
ds^2 = g_{\mu\nu}\, (dx^\mu+N^\mu\, dw) (dx^\nu+N^\nu\, dw) - (N\, dw)^2,
\label{ADMg}
\eeq
where $g_{\mu\nu}$ is the 4D metric of a hypersurface at fixed $w$ and from Eq.~(\ref{5Dmetric}) the lapse $N$ and shift $N^\mu$ functions are given by,

\beq
N = 1+\frac{\delta g_{44}}{2}, \ \ \ \ \ N_\mu = (F,\nabla C),
\label{lapseshift}
\eeq
where we have used linear perturbation theory for $N$. It is also useful to write the extrinsic curvature tensor of a hypersurface at fixed $w$,

\beq
K_{\mu \nu} = \frac{1}{2N} \Big[ N_{\mu;\nu}+ N_{\nu;\mu}- \partial_w g_{\mu\nu} \Big],
\label{extrinsic}
\eeq
where the covariant derivatives are with respect to the 4D metric $g_{\mu\nu}$. From this we have, for the background metric (bars denote background values)

\beq
\bar{K}_{00} = -{1\over 2}({\rm e}^{2\beta})', \ \ \ \ \ \bar{K}_{ij} ={1\over 2}({\rm e}^{2\alpha} )'\, \delta_{ij},
\eeq
where primes denote derivatives in the extra spatial dimension. Therefore, the extrinsic curvature scalar reads

\beq
\bar{K}=- (3\alpha' + \beta'),
\eeq
whereas for perturbations we have to linear order

\beq
\delta K = {\rm e}^{-2\beta} \Big[ F  (3\dot{\alpha}-\dot{\beta})+ \dot{F} \Big] + (3\psi'-\phi') - {\rm e}^{-2\alpha} \nabla^2 C .
\label{dK}
\eeq
A similar (though longer) expression holds for the full extrinsic curvature tensor. This will imply (after we solve the bulk equations and develop the quasistatic solution) that the extrinsic curvature will be dominated by second derivatives of the brane-bending mode at subhorizon scales, and by normal derivatives at larger (but still smaller than Hubble) scales. At scales larger than the Hubble radius the contributions from $F$ become important, but we won't deal with such wavelengths in this paper, thus $F$ will play essentially no role. Since extrinsic curvature is the new degree of freedom that makes gravity different from GR, this shows which metric perturbations are going be relevant for structure formation.

\section{Field Equations and Linear Theory}

\subsection{Basics}

The field equations corresponding to brane-induced gravity in five dimensions are~\cite{2000PhLB..484..112D},

\beq
\frac{\sqrt{g^{(5)}}}{2r_c}G^{(5)}_{AB} + G_{AB} \, [\dD]= 8\pi G \ T_{AB} \  [\dD],
\label{EinsteinDGP}
\eeq
where $G^{(5)}_{AB}$ is the 5D Einstein tensor, and $G_{AB},T_{AB}$ the standard $3+1$ Einstein and stress-energy tensors, which vanish when any index is larger than $3$; this condition for $T_{AB}$ enforces there is no flux of energy-momentum to the extra dimension (recall the position of the brane is fixed at $w=0$ even in the presence of fluctuations). In Eq.~(\ref{EinsteinDGP}), $g^{(5)}$ denotes the absolute value of the 5D metric determinant, and $[\dD] \equiv \dD(w) \sqrt{g^{(4)}}$. The parameter $r_c$ denotes the cross-over scale from 4D to 5D behavior, indeed one can rewrite by dimensional analysis Eq.~(\ref{EinsteinDGP}) for a point source at the brane as

\beq
\frac{1}{r_c}\, \frac{\phi}{r^2} + \frac{\phi}{r^2}\, \frac{1}{r} \sim \frac{1}{r^3}\, \frac{1}{r},
\label{DimAn}
\eeq
where $\phi$ denotes a generic component of the metric. Therefore, for $r\ll r_c$ one has the standard 4D behavior $\phi \sim r^{-1}$, whereas for $r \gg r_c$ the 5D behavior $\phi \sim r^{-2}$ is obtained. More precisely,  small perturbations about Minkowski  $g_{\mu\nu}=\eta_{\mu\nu}+h_{\mu\nu}$ ($h_{\mu\nu} \ll 1$) the metric at the brane can be readily obtained in momentum space~\cite{2002PhRvD..65b4031D}

\beqa
\tilde{h}_{\mu\nu} &=& G(p)\ (\tilde{T}_{\mu\nu}-{1\over 3}\eta_{\mu\nu}\, \tilde{T}) \nonumber \\
&=& G(p)\ (\tilde{T}_{\mu\nu}-{1\over 2}\eta_{\mu\nu}\, \tilde{T}) +{1\over 6} G(p)\ \eta_{\mu\nu}\, \tilde{T}, \nonumber \\ & & 
\label{Solvehmunu}
\eeqa
where a tilde denotes objects in Fourier space (depending on four-momentum $p^\mu$), $T\equiv T^\mu_\mu$, and 
\beq
G(p) = {8\pi G \over p^2+p/r_c},
\label{GMink}
\eeq
from which it follows Eq.~(\ref{DimAn}) for a static point source. We will come back to Eq.~(\ref{GMink}) when we discuss linear cosmological perturbations as the unusual $p^{-1}$ dependence will lead to interesting scale dependence of the growth factor at large scales. 
In the second line of Eq.~(\ref{Solvehmunu}), we have decomposed the tensor structure of the full theory (previous line) in terms of the standard massless spin-2 part (first term) and the helicity-zero part (second term) that effectively acts as a scalar. The fact that the full theory in the massless limit $pr_c \gg 1$ does not recover the massless case (GR) is the well-known van Dam-Veltman-Zakharov  (vDVZ) discontinuity~\cite{1970PhRvD...2.2255I,1970NuPhB..22..397V,1970JETPL..12..312Z}. As conjectured by Vainshtein in massive gravity~\cite{1972PhLB...39..393V}, the discontinuity should disappear in the full nonlinear theory (in which the helicity-zero mode gets suppressed), otherwise one could rule out a massive graviton ($m\sim r_c^{-1}$ here) from local observations no matter how tiny its mass. Understanding this Vainshtein mechanism for realistic cosmological perturbations is one of the main goals of this paper. We shall see that the behavior of the graviton propagator just discussed has a close analogy to the modified Poisson propagator for cosmological perturbations, as expected.

It will be useful for future reference to write the gravitational part of the action for Eq.~(\ref{EinsteinDGP}) in ADM form,

\beqa
S_{\rm grav} &=& \frac{1}{16\pi G} \int d^4x\, dw \, \sqrt{g^{(5)}} \, \Big[ R \ \delta_{\rm D}(w)  \nonumber \\ & & 
+\frac{1}{2r_c} N  (R + K^2 -K_{\mu\nu}K^{\mu\nu})\Big],
\label{Sgrav}
 \eeqa

where $R$ is the Ricci scalar for the 4D metric. The first term is the usual Einstein-Hilbert term, whereas the second contribution proportional to the lapse function $N$ is the 5D Einstein-Hilbert term decomposed in ADM form into intrinsic ($R$) and extrinsic ($K$) curvature of hypersurfaces at constant $w$. For the Ricci scalar of the background we have,

\beq
\bar{R}= -6  {\rm e}^{-2\beta} (2\dot{\alpha}^2-\dot{\alpha}\dot{\beta}+\ddot{\alpha})
\eeq

whereas for perturbations,

\beqa
\delta R &=& 2  {\rm e}^{-2\alpha} \nabla^2(\phi-2\psi) - 2 \bar{R}\, \phi \nonumber \\ & &  
+ 6  {\rm e}^{-2\beta}  
(\dot{\alpha}\dot{\phi}+4 \dot{\alpha}\dot{\psi}-\dot{\beta}\dot{\psi}+\ddot{\psi})
\label{pertR}
\eeqa

The stress-energy tensor we assume throughout this paper is that corresponding to a dark matter fluid, 

\beq
T^{\mu\nu}=\bar{\rho}\, (1+\delta)\, u^\mu u^\nu,
\label{Tmunu}
\eeq
where $\bar{\rho}$ is the mean density, $\delta$ describes density fluctuations and the four-velocity,

\beq
u^\mu= \frac{v^\mu}{\sqrt{g_{\mu\nu}v^\mu v^\nu}},
\eeq
where $v^\mu=d x^\mu/d\tau=(a,v^i)$ with $\v$ the standard velocity fluctuations about the Hubble flow, and $\tau$ is the conformal time. The only non-vanishing component of the stress tensor fluctuations are (in linear perturbation theory),

\beq
\delta T^0_0=\bar{\rho}\, \delta,
\eeq
\beq
\delta T^0_i=- a \bar{\rho}\, v_i.
\eeq

From this, conservation of energy and momentum gives respectively (always to first order),

\beq
\label{DivT0}
\dot{\delta}-3\dot{\psi}+\frac{1}{a}\theta=0,
\eeq
\beq
\label{DivTi}
\dot{\theta}+H\theta+\frac{1}{a}\nabla^2\phi=0.
\eeq
The first is the standard continuity equation with the extra term representing the change in volume as the universe expands due to the fact that $a(1-\psi)$ is the perturbed scale factor. The second is the geodesic equation for the fluid, in the approximation of weak fields and small velocities. These equations are independent of the theory of gravity (as long as it is a metric theory and stress energy is conserved), and must be supplemented by the field equations of the gravity theory under consideration, which relates stress energy fluctuations to metric fluctuations.

\subsection{Background Evolution}

Let's briefly recall the evolution of the background in the self-accelerated branch, this will also be useful to set the time variables we are going to use when discussing perturbations. From the $00$ field equation we obtain the modified Friedmann equation,

\beq
H^2= \frac{H}{r_c}+\frac{8}{3}\pi G \bar{\rho},
\label{FeqDGP}
\eeq
Since $ \bar{\rho}>0$, the theory makes sense only when $r_cH>1$, thus probing distances larger than the crossover scale $r_c$ necessarily implies considering perturbations of wavelength larger than the Hubble radius. As time goes on, one immediately sees from Eq.~(\ref{FeqDGP}) that since $ \bar{\rho}\rightarrow0$, one enters a de~Sitter phase where the universe expands exponentially with $H=1/r_c$. The acceleration follows from the $ii$ field equation, which gives 
\beq
\dot{H}=-\frac{3}{2}H^2\, \frac{r_cH-1}{r_cH-1/2},
\label{hdot}
\eeq
which implies that the acceleration of the Universe is given by

\beq
\frac{\ddot{a}}{a}=\dot{H}+H^2=-H^2\ \frac{\eta-2}{2\eta-1},
\eeq
where we defined,
\beq
\eta \equiv r_cH,
\eeq
thus the Universe starts to accelerate when  $H$ drops below the critical value $\eta_{\rm crit}=2$. In the early universe $\eta\gg 1$, in fact $\eta$ can be thought of as a time variable which decreases with cosmic time ($\eta>1$, achieving $\eta=1$ in the infinite future when the dark matter density asymptotes to zero). In terms of this time variable, the scale factor reads,
\beq
a(\eta)=\Big[\frac{\eta_0(\eta_0-1)}{\eta(\eta-1)}\Big]^{1/3},
\label{aeta}
\eeq
and $\eta_0=H_0r_c$ can also be written in terms of the $z=0$ matter density, $\eta_0=(1-\Omega_m^0)^{-1}$. One can go back and forth between $a$ and $\eta$ using Eq.~(\ref{aeta}) and its inverse relation, $2\eta=1+\sqrt{1+4\eta_0(\eta_0-1)a^{-3}}$.

We can define the {\em acceleration} parameter,

\beq
q \equiv \frac{\ddot{a}a}{\dot{a}^2} = \frac{2-\eta}{2\eta-1},
\eeq

which goes from $-1/2$ in the early universe to $1$ as the universe gets into the purely de Sitter phase. We can then write simply,

\beq
 {\rm e}^{\alpha} = a\, (1+|w| H), \ \ \ \ \  {\rm e}^{\beta} =  1+|w|Hq .
 \label{albeHq}
 \eeq

Finally, note that before the universe enters into the acceleration phase ($\eta>2$), $q<0$ and from Eqs.~(\ref{5Dmetric}) and (\ref{albeHq}) it follows that there is a Rindler horizon in the bulk where $\beta \rightarrow-\infty$ located at

\beq
w_{h}= - (qH)^{-1},
\label{Rindler}
\eeq
which becomes $2H^{-1}$ in the early universe, and goes to infinity when the universe approaches the accelerated phase,  and disappears afterwards.

In what follows we will use $\eta$ and $q$ to characterize the time dependences of the scale factor in this model.

\subsection{Linear Perturbations: at the Brane}

We now consider linear metric perturbations at the brane.  Since the dark matter has negligible anisotropic stress, $\delta T^i_j=0$, and thus from the  $ij$-equations ($i\neq j$) at $w=0$ one obtains the following simple boundary condition for the brane-bending mode $C$,
\beq
C_0=r_c\ (\psi_0-\phi_0),
\label{CBC}
\eeq
where a $0$ subscript denotes quantities evaluated at the brane. The $00$-equation at the brane gives the modified Poisson equation in linear theory,

\begin{widetext}
\beq
\nabla^2\psi_0 - \frac{1}{2r_c}\nabla^2C_0 + \frac{3a^2}{2r_c} \Big( \psi_0' +H F_0 \Big) =
3a^2H\Big \{ \dot{\psi_0}+H\phi_0 \Big\} + \, 4\pi G a^2 \bar{\rho}\, \delta. 
\label{dg00}
\eeq
\end{widetext}
At this point it is worth making a few observations.  First, note that we can obtain the GR limit by taking $\psi_0=\phi_0$ in Eq.~(\ref{CBC}) which sets to zero $C_0$ in Eq.~(\ref{dg00}), and then ignoring the contribution in parenthesis (effectively setting $r_c$ to infinity) gives the Poisson equation in GR. The terms in braces on the right hand side of Eq.~(\ref{dg00}) correspond to the usual contributions in GR from retardation effects that dominate at superhorizon scales, while at subhorizon scales they are negligible compared to $\nabla^2\psi_0$. The reason that normal derivatives only appear at most to first order in Eq.~(\ref{dg00}) is that at least another first normal derivative of the background metric (that has a jump at $w=0$) is needed to get a nonzero contribution localized in an infinitesimal brane.

We are interested in finding the growth of structure at subhorizon scales where time derivatives are small compared to spatial gradients, and thus one can assume a quasistatic evolution. In this case, Eq.~(\ref{dg00}) becomes,

\beq
\frac{1}{2}\, \nabla^2(\phi_0+\psi_0) =  4\pi G a^2 \bar{\rho}\, \delta,
\label{PoissonTrucho}
\eeq
where we have used the constraint in Eq.~(\ref{CBC}) to replace the brane-bending mode $C_0$ in terms of potentials $\psi_0,\phi_0$, and have neglected the terms in parenthesis and braces as they involve spatial derivatives of order less than second supressed by the Hubble radius or $r_c\sim H_0^{-1}$. Then, all that remains to find is the relation between $\phi_0$ and $\psi_0$. This can be obtained directly from the $44$ bulk equation close to the brane ($w=0^+$), which in the same approximation reads,
\beq
 \nabla^2(\psi_0-\phi_0)= \frac{1}{\eta(q+2)}\, \nabla^2(2\psi_0-\phi_0),
 \label{dg44Trucho} 
\eeq
and thus the system is closed. This is essentially the argument given in~\cite{2004PhRvD..69l4015L} to derive the growth factor at subhorizon scales. Note that Eq.~(\ref{dg44Trucho}) relates the extrinsic curvature, Eq.~(\ref{dK}), to the Ricci scalar, Eq.~(\ref{pertR}), at subhorizon scales; we shall come back to this below. At high redshift,  $\eta=Hr_c\gg 1$ while $q\rightarrow-1/2$, and Eq.~(\ref{dg44Trucho}) sets $\psi_0=\phi_0$ and GR is recovered in Eq.~(\ref{PoissonTrucho}).  

The modified Poisson equation that results from Eqs.(\ref{PoissonTrucho})-(\ref{dg44Trucho}) is,

\beq
\nabla^2\phi_0 = 4\pi G_{\rm eff} a^2 \bar{\rho}\, \delta,\ \ \ \ \ 
G_{\rm eff}\equiv G\  \frac{2\eta(q+2)-4}{2\eta(q+2)-3},
\label{modPoisson}
\eeq
thus modification of gravity in this approximation can be understood as a time-dependent (but scale-independent) modification of the gravitational constant. Note that at late times, in the de Sitter limit, $\eta,q \rightarrow 1$ and $G_{\rm eff} \rightarrow (2/3) G$. Thus gravity becomes weaker leading to a slower growth of structure compared to a dark energy model with the same expansion history.

These arguments, while certainly reasonable, are not rigorous enough to understand the limit of applicability of the approximations made. For example, that $G_{\rm eff} $ is only a function of time results from keeping only the highest (second) spatial derivatives. We would like to understand the limitations of these results at both large and small scales. The limitations at large scales come from neglecting the normal derivatives to the brane, $\psi_0'$, for which we need to solve the equations in the bulk~\cite{2006JCAP...01..016K}. The limitations at small scales come from the fact that brane-induced gravity becomes GR at small scales through the Vainshtein mechanism, where nonlinear effects suppress the brane-bending mode $C_0$. For this we need to understand the nonlinearities responsible for this mechanism and study at what scale they do operate. We first consider the linear bulk equations, postponing until section~\ref{NL} the discussion of nonlinear effects.

\subsection{Linear Perturbations: in the Bulk}
\label{LPbulk}

Before we discuss the solution of the bulk equations, it will be useful to split perturbations into two classes according to their behavior under gauge transformations that shift the brane position, $w \to w + \delta x^4$. We define two new metric variables, 

\beq
\Phi \equiv \phi+\psi +(\beta'-\alpha')\, C, \ \ \ \ \ f \equiv F -\dot{C},
\label{PhifDef}
\eeq

which, as can be seen from Eqs.~(\ref{phiGT}-\ref{dg44GT}), are invariant under shifts of the brane position. We also define a new metric variable,

\beq
\Psi \equiv 2\psi-\phi,
\label{PsiDef}
\eeq

which gives the leading contribution to the perturbed Ricci scalar, Eq.~(\ref{pertR}), at subhorizon scales. We keep the brane bending mode as our fourth metric variable, thus we are looking for solutions in the bulk for $\Phi,f,\Psi,C$, where the first two variables are invariant under gauge transformations involving only $w$, and the other two are not. As we shall see, this split between $\Phi, f$ and $\Psi,C$ will play a key role in solving the bulk equations.

The equations that determine the shift vector $N_\mu$ in the bulk are very simple. The $ij$ field equation reads,

\beq
3 C' + 2(2\alpha'+\beta')\, C = 2\Psi-\Phi, 
\label{ijbulk}
\eeq
which determines $C$ once $\Phi$ and $\Psi$ are known. After using this result, the $0j$ field equations give 
\beq
f' + (\alpha'+\beta')\, f = 2 (\dot{\Phi}+\dot{\alpha}\, \Phi).
\label{0ibulk}
\eeq
This determines $f$ from the evolution of $\Phi$.  An equation for $\Phi$ follows from subtracting one third of the $ii$ 
field equations from the $00$ equations,

\beq
\Box_5\, \Phi = 4 \dot{\alpha}  {\rm e}^{-2\beta} \dot{\Phi} + 4(\alpha'-\beta') \Psi' + \{\Phi,\Psi,f,C \},
\label{PhiBulk}
\eeq
where the 5D d'Alembert operator in the background metric reads
\beq
\Box_5 \equiv \Box -(3\alpha'+\beta')\, \partial_w-\partial^2_w,
\label{Box5}
\eeq
and the 4D one,
\beq
\Box \equiv {\rm e}^{-2\beta}\, \partial^2_t + (3\dot{\alpha}-\dot{\beta})\, {\rm e}^{-2\beta} \,\partial_t - {\rm e}^{-2\alpha} \nabla^2,
\label{Box4}
\eeq
and the term in braces in Eq.~(\ref{PhiBulk}) denotes terms that involve the four metric variables (but none of their derivatives).  All the terms that do not involve $\Phi$ have the property that their coefficients vanish in the de~Sitter limit ($q=1$), as can be seen explicitly for the $\Psi'$ term, i.e. $\alpha'=\beta'=H/(1+wH)$ in the de~Sitter phase. In this limit, $\{\}= 2\dot{\alpha}^2\, {\rm e}^{-2\beta} \,\Phi$, and Eq.~(\ref{PhiBulk}) becomes a closed equation for $\Phi$. This shows that during the accelerated phase we are very close to having decoupled the bulk equations, even beyond the quasistatic approximation.

So far we have written exact bulk equations in linear theory. The remaining bulk equations are in some cases rather complicated but we are interested in sub-horizon perturbations where time derivatives are suppressed compared to spatial gradients. Therefore, 
in the following we make the approximation that we consider modes inside the Hubble radius, i.e. $(k/aH)^2 \gg 1$, where $k$ is the comoving wavenumber. This is a very good approximation since, e.g.  at $z=0$
\beq
 \frac{k}{aH} = 300~\Big(\frac{k}{0.1 \kvecMpc}\Big).
\label{eps}
\eeq

In this approximation, Eq.~(\ref{PhiBulk}) becomes

\beq
\Phi'' + (3\alpha'+\beta') \Phi' + {\rm e}^{-2\alpha} \nabla^2 \Phi 
\approx 0,
\label{PhiBulkQS}
\eeq
where we have also neglected $\Psi'$, $\Psi$, $f$ and $C$ terms (all of which disappear as the universe approaches the de Sitter phase). This will be further justified shortly. Equation~(\ref{PhiBulkQS}) can be readily solved in Fourier space as 

\beq
\Phi = \Phi_0 \ (1+s)^{-k/aH}\ \ \ \ \ \ \ \ \, s \equiv wH
\label{QSsol}
\eeq
where we have dropped the solution that diverges at infinity. Strictly speaking, before the universe accelerates one must impose vanishing boundary conditions at the Rindler horizon given by Eq.~(\ref{Rindler}), but the quasistatic approximation breaks down at such large distances away from the brane, and in any case Eq.~(\ref{QSsol}) decays so fast that any mixing with the second solution is going to be negligibly small. Equation~(\ref{QSsol}) is analogous to the quasistatic bulk solution obtained  in~\cite{2006JCAP...01..016K} for the master variable (however, it is not the same as the master variable is a gauge invariant quantity, and $\Phi$ is only invariant under gauge transformations involving only the extra-dimension coordinate). 

We can check the validity of the quasistatic solution by comparing the time derivatives to the other terms in the Eq.~(\ref{PhiBulk}). We have

\beq
\ddot{\Phi} \simeq \frac{k^2}{a^2}\ \Big[(1-q)\frac{s}{1+s} +q\, \ln(1+s)\Big]^2\ \Phi,
\label{Phiddot}
\eeq
and compared to spatial derivatives in the left hand side of Eq.~(\ref{PhiBulk}) we see that time derivatives become important only far from the brane, i.e. when $s\simeq 1$ or $w\simeq H^{-1}$; at this point the quasistatic solution in Eq.~(\ref{QSsol}) has decayed exponentially to zero for the wavenumbers of interest, see Eq.~(\ref{eps}). 

Having a solution for $\Phi$ we can now solve for $f$ from Eq.~(\ref{0ibulk}). Here one must be careful to include time derivatives that act as a source of $f$ in Eq.~(\ref{0ibulk}), leading to

\beq
f \simeq -2 \Big[(1-q)s+q(1+s)\ln(1+s)\Big] \Phi= - 2{\rm e}^{2\alpha}\nabla^{-2}\dot{\Phi}',
\label{fsol}
\eeq
where we have set the homogeneous solution $f_{\rm H}\propto (1+s)^{-1}(1+qs)^{-1}$ to zero since it violates the $i4$ bulk equations, as we shall see in Eq.~(\ref{i4bulk2}) below. Note that close to the brane a subleading term of order $\Phi\, (aH/k)$ becomes dominant since Eq.~(\ref{fsol}) vanishes as $s\rightarrow 0$. At the brane ($s=0^+$) we then have

\beq
f_0 \approx -2 \Phi_0\ \Big( \frac{aH}{k}\Big)\  \Big(2+ \frac{d\ln \Phi_0}{d\ln a} \Big).
\eeq
In what follows we neglect this subleading contribution, and use Eq.~(\ref{fsol}) everywhere in the bulk. Then $f$ is at most of order $\Phi$. Note that neglecting $f$ altogether results in a shift vector that is a pure four-gradient, i.e. $N_\mu = \nabla_\mu C$. 

It remains to find the dynamics of $\Psi$ and of the brane-bending mode $C$ which determine, respectively, the Ricci  scalar  and the extrinsic curvature in constant $w$ hypersurfaces at subhorizon scales. As it is clear from the ADM formalism, Eqs.~(\ref{lapseshift}) and~(\ref{Sgrav}),  the 44 field equation relates the Ricci scalar to the extrinsic curvature (the Gauss-Codazzi equation),
\beq
R=  K_{\mu\nu}K^{\mu\nu}-K^2,
\label{Eq44}
\eeq
which to linear order gives, in the subhorizon approximation (using the same notation as in Eq.~\ref{PhiBulk})

\begin{widetext}

\beq
(2\alpha'+\beta')\Box C -\Box \Psi +(\beta'-3\alpha')\Psi'= {\rm e}^{-2\beta}\, \ddot{\Phi} -\beta'\Phi'+\{f,C,\dot{C},\Phi,\dot{\Phi},\Psi\}.
\label{Eq44lin}
\eeq

\end{widetext}
Again, working in the quasistatic approximation, and neglecting $\Psi'$ compared to $\nabla^2\Psi$, which will be justified shortly, we obtain a relationship between $\Psi$ and $C$ that can be written,
\beq
\Psi = (2\alpha'+\beta')C+{\rm e}^{2(\alpha-\beta)}\nabla^{-2}\ddot{\Phi}-\beta'{\rm e}^{2\alpha}\nabla^{-2}\Phi'.
\label{PsiClin}
\eeq
In principle one could add to this the solution of $\Box \Psi =(\beta'-3\alpha')\Psi'$, which is $\Psi=(1+s)^{-\bar{k}^2/3} (1+2qs/(3-q))^{-\bar{k}^2/6}$, with $\bar{k}=k/aH$, but such term would violate the $i4$ field equation below. This is important because otherwise the solution for the linear growth factor would be different at subhorizon scales. Replacing Eq.~(\ref{PsiClin}) in Eq.~(\ref{ijbulk}) and ignoring subleading terms gives a very simple equation for $C$ in terms of $\Phi$ alone,
\beq
3C'=-\Phi,
\label{CPhi}
\eeq
which leads to
\beq
C=C_0 +\frac{\Phi_0}{3}\Big(\frac{a}{k}\Big) \Big[(1+s)^{-k/aH}-1\Big],
\label{Cbulklin}
\eeq
and this in turn gives a solution for $\Psi$ from Eq.~(\ref{PsiClin}).

We are now in a position to check the validity of the approximations made. First, to go from Eq.~(\ref{PhiBulk}) to Eq.~(\ref{PhiBulkQS}) we have dropped the terms involving $f$, $\Psi$, $\Psi'$ and $C$, apart from assuming the quasistatic approximation for $\Phi$. Since $f$ is at most of order $\Phi$, Eq.~(\ref{fsol}), it is justified to drop it compared to $\nabla^2\Phi$. The remaining terms involving $\Psi$, $\Psi'$ and $C$ are potentially dangerous, as $C$ does not decay into the bulk and $\Psi$ only decays slowly, thus they can become larger than $\nabla^2\Phi$ as it becomes strongly damped in the bulk. However, using Eqs.~(\ref{PsiClin})-(\ref{Cbulklin}) it follows that all these terms in Eq.~(\ref{PhiBulk}) cancel leaving only contributions proportional to $\Phi$. This is as expected from the properties of fields under gauge transformations that shift the brane position. Finally, we see from Eqs.~(\ref{PsiClin})-(\ref{Cbulklin}) that it is also justified to drop $\Psi'$ compared to $\nabla^2\Psi$ in the quasistatic regime.

Other checks can be done from the remaining bulk equations. The $i4$ field equation reads, 

\beqa
\dot{f}+(\dot{\alpha}-\dot{\beta})f &=& -\frac{2}{3}\, {\rm e}^{2\beta}\Big[ 
3 \Psi'+(\alpha'-\beta')(2\Phi-\Psi) \nonumber \\
& & + (8\alpha'^2-\alpha'\beta'+2\beta'^2)\, C
\Big]
\label{i4bulk}
\eeqa
which in view of Eqs.~(\ref{PsiClin}-\ref{Cbulklin}) can be written as
\beq
\dot{f}+(\dot{\alpha}-\dot{\beta})f = 2\beta' {\rm e}^{2\beta}\Phi - 2 {\rm e}^{2\alpha} \nabla^{-2}\ddot{\Phi}',
\label{i4bulk2}
\eeq

which is solved by Eq.~(\ref{fsol}), and does not support a slow decay into the bulk, i.e. the homogenous solution of Eq.~(\ref{0ibulk}), $f_{\rm H}\propto (1+s)^{-1}(1+qs)^{-1}$, does not solve Eq.~(\ref{i4bulk2}) with zero right hand side, and thus must be discarded. Finally, the $04$ bulk equation can be shown to hold by using Eq.~(\ref{fsol}) for $f$ as well.

\begin{figure}[t!]
\begin{center}
\includegraphics[width=0.5\textwidth]{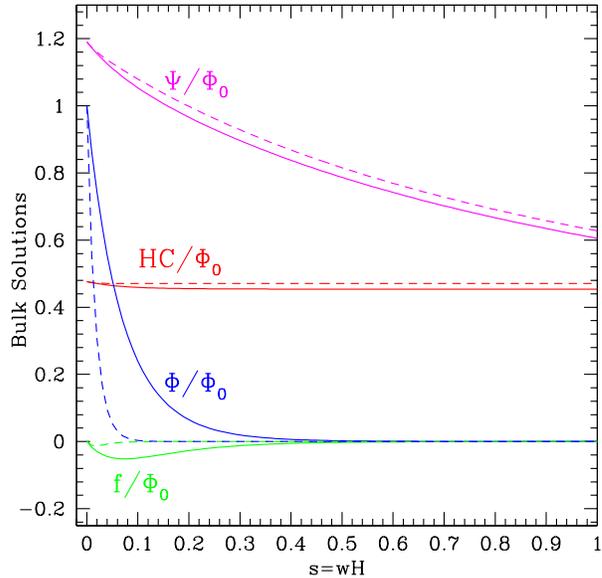}
\caption{Bulk solutions, for $k=0.005\kvecMpc$ (solid) and $k=0.02\kvecMpc$ (dashed), as a function of $s=wH$ for redshift $z=0$, when the matter density is given by $\Omega_m^0=0.2$. The brane is located at $s=0$. All metric variables are normalized by the value of $\Phi$ at the brane, $\Phi_0$. Fields invariant under shifts of the brane position ($\Phi,f$) decay strongly into the bulk, whereas $C,\Psi$ do not.}
\label{BulkSolFIG}
\end{center}
\end{figure}

These results are summarized in Fig.~\ref{BulkSolFIG}, where we show the four bulk solutions as a function of the coordinate into the bulk in units of the Hubble radius, $s\equiv wH$, for two different wavectors $k=0.005\kvecMpc$ (solid) and $k=0.02\kvecMpc$ (dashed). Note that fields invariant under shifts of the brane position ($\Phi,f$) are concentrated at the brane and leak only partially into the bulk, whereas those who are not ($C,\Psi$) approach a constant far from the brane. However, one should keep in mind that this neat separation of bulk behavior according to symmetry breaks down at small scales (high-$k$) when nonlinearities are included, as we shall discuss in section~\ref{NL}.

\subsection{The Linear Quasistatic Solution}

Now we can go back to the brane equations and finalize the linear quasistatic solution. From Eqs.~(\ref{PhifDef})-(\ref{PsiDef}) we can map back to the $\phi.\psi,F$ variabls relevant to the Poisson equation, Eq.~(\ref{dg00}),

\beqa
\label{phimap}
\phi &=& \frac{1}{3}\Big[2(\alpha'-\beta')C+2\Phi-\Psi\Big],\\
\psi &=& \frac{1}{3}\Big[(\alpha'-\beta')C+\Phi+\Psi\Big],
\label{psimap}
\eeqa
from which we can obtain the normal derivatives at the brane,

\beq
\phi_0'\simeq 2\psi_0'\simeq \frac{2}{3}\Phi_0'=-\frac{2}{3}\frac{k}{a}\Phi_0,
\label{normder}
\eeq
plus subleading (order $H\Phi_0$) contributions. We thus see that normal derivatives are suppressed compared to the leading contributions in the Poisson equation Eq.~(\ref{dg00}) by one inverse power of  $kr_c/a$. These terms are genuinely of quasistatic nature (i.e. they give rise to the 5D decay of the gravitational potential for static sources), and thus we must keep them as part of the quasistatic solution. They are dominant compared to retardation effects standard from GR (which are also modified here at late times by 5D dynamics), which are suppressed by {\em two} inverse powers of $k/aH$. Clearly, $kr_c/a$ compares a physical scale $a/k$ against the spatial scale $r_c$, whereas $k/aH$ compares the light crossing time across a perturbation of size $a/k$ to the Hubble time $H^{-1}$, a measure of retardation effects.

An important feature of brane-induced gravity is that from the point of view of standard 4D physics, the bulk dynamics introduces through normal derivatives such as those in Eq.~(\ref{dg00}), unusual non-local operators. From Eq.~(\ref{normder}) we have that normal derivatives of $\Phi_0$ can be written in configuration space as

\beq
\Phi_0' = -\frac{1}{a}\sqrt{-\nabla^2}\ \Phi_0.
\label{sqrtnabla}
\eeq
This is a consequence of the 5D nature of the gravity theory.

We can then summarize the linear quasistatic equations as follows. To make equations compact, we will first make use of different (though linearly related) metric variables. We later write down everything in terms of  the Newtonian potential $\phi_0$. Using Eq.~(\ref{PsiClin}) at $w=0^+$ we have,
\beq
H(q+2)\, \nabla^2  C_0 +Hqa\sqrt{-\nabla^2} \, \Phi_0 = \nabla^2 \Psi_0,
\label{44lin0}
\eeq
which is a more accurate version of Eq.~(\ref{dg44Trucho}). To write the Poisson equation, Eq.~(\ref{dg00}), in the quasistatic approximation, we can drop the terms in braces in Eq.~(\ref{dg00}) (as usual in GR) since they are down from the leading terms by the factor $(aH/k)^2$, {\em and} also we can drop $F_0$ as from the bulk solution $F_0 \simeq \dot{C}_0$, and thus this term is subdominant by the same factor to the term with $\nabla^2C_0$. Then we have,
\beq
\nabla^2\psi_0-\frac{1}{2r_c}\nabla^2C_0 -\frac{a}{2r_c}\sqrt{-\nabla^2}\,\Phi_0 = 4\pi Ga^2\bar{\rho}\, \delta.
\label{PoissonQSprelim}
\eeq

Note, as argued above, that the nonlocal terms in these equations belong to the quasistatic approximation. For a point source of mass $m$, $\delta \sim 1/r^3$, and the right hand side of the Poisson equation scales as $r_g/r^3$, where $r_g\propto Gm$ is the gravitational (Schwarzschild) radius of the source. The first two terms in Eq.~(\ref{PoissonQSprelim}) scale as $\phi/r^2$, where $\phi$ is the order of magnitude of metric perturbations, while the third goes as $\phi/(rr_c)$, then we can symbolically write the Poisson equation as (compare to Eq.~\ref{DimAn}),
\beq
\frac{\phi}{r^2} + \frac{\phi}{rr_c} \sim \frac{r_g}{r^3},
\label{to5D}
\eeq
and thus the normal derivative term governs the transition of the static on-brane potential from the usual 4D behavior $\phi \sim r_g/r$ for $r\la r_c$ to the 5D behavior $\phi \sim (r_gr_c)/r^2$ for $r\ga r_c$. Eq.~(\ref{PoissonQSprelim}) thus reflects precisely the behavior of the propagator in Eq.~(\ref{GMink}) for static sources, with the nonlocality in the Poisson equation a consequence of the $p^{-1}$ behavior of the on-brane propagator. This is a feature of extra-dimensional theories of gravity that cannot take place in a local model of dark energy (no matter how complicated), and it is interesting to note that it is expected to happen in theories with more than one extra dimension as well, as long as there is a 5D regime before hitting the truly large distance regime (as in ``cascading" models, see~\cite{2007JHEP...05..045K,2008JCAP...02..011D,2008PhRvL.100y1603D}).

As we shall see, these non-local corrections become interesting well before scales of order $r_c$ and improve the validity of the quasistatic approximation at large scales compared to the quasistatic solution of~\cite{2004PhRvD..69l4015L,2006JCAP...01..016K} which sets the normal derivatives to zero. Therefore, it is expected that at least some of the deviations reported  at large scales in numerical solutions that include normal derivatives~\cite{2007PhRvD..75f4002S,2008PhRvD..77h3512C} can be explained by this simple result.

Note that in Eqs.~(\ref{44lin0})-(\ref{PoissonQSprelim}), the brane metric variables are all related through Eqs.~(\ref{CBC}),~(\ref{PhifDef}) and~(\ref{PsiDef}). Thus there is only one independent brane potential that obeys a Poisson-like equation. We want to compute the growth of density perturbations, and since the Newtonian potential $\phi_0$ is responsible for accelerations, Eq.~(\ref{DivTi}), we write everything in terms of the Newtonian potential. Thus we have the effective Poisson equation in Fourier space

\begin{widetext}
\beq
-k^2 \phi_0 \ 
\Big\{\frac{2\eta(2+q)-3}{2\eta(2+q)-4}+\Big(\frac{3a}{kr_c}\Big) \Big(\frac{\eta(1+q)-1}{\eta(2+q)-2}\Big)^2 \Big\} 
= 4\pi G a^2 \bar{\rho}\, \delta,\ \ \ \ \  {\rm or}\ \ \ -k^2 \phi_0 \equiv 4\pi G_{\rm eff}(k,\eta) a^2 \bar{\rho}\, \delta,
\label{PoissonQSlin}
\eeq
where we have defined an effective gravitational constant $G_{\rm eff}$ that depends on scale and time~\footnote{Note that our definition of $G_{\rm eff}$ differs from that in~\cite{2008PhRvD..78f3503J}, which uses $-k^2 \psi_0 \equiv 4\pi G_{\rm eff}(k,\eta) a^2 \bar{\rho}\, \delta$. }. 
For completeness, we note that one can also derive the following Poisson-like equation for the brane-bending mode,

\beq
 -k^2\, {C_0 \over r_c} \ 
 \Big\{[2\eta(2+q)-3]  -\Big(\frac{2a}{kr_c}\Big)(2\eta q-1)[\eta(2+q)-2] \Big\}  = 8\pi G a^2 \bar{\rho}\, \delta.
 \label{BBMQSlin}
 \eeq
 \end{widetext}

\begin{figure}[t!]
\begin{center}
\includegraphics[width=0.5\textwidth]{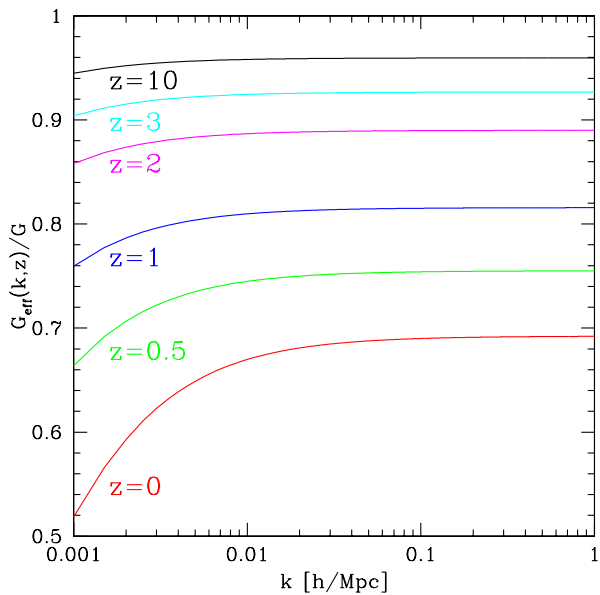}
\caption{The effective gravitational constant $G_{\rm eff}(k,z)$ that governs the linearized Poisson equation for the Newtonian potential, Eq.~(\protect\ref{PoissonQSlin}), as a function of $k$ for different redshifts $z=0,0.5,1,2,3,10$ (from bottom to top). Nonlinear effects modify significantly this behavior at small scales, see Fig.~\protect\ref{GeffNL} below.}
\label{GeffFIG}
\end{center}
\end{figure}

Figure~\ref{GeffFIG} shows $G_{\rm eff}(k,z)$ as a function of $k$ for different redshifts. We see that at all times $G_{\rm eff}$ is less than $G$, more so at late times (when the modifications away from GR become more important, as the universe accelerates), and at large scales where the transition to the 5D regime of even weaker gravity takes place, inducing scale dependence. For the purpose of this plot we have assumed $\Omega_m^0=0.2$ for which $\eta_0=1.25$ and the universe starts accelerating at $z=0.86$. 

The two features seen in Fig.~\ref{GeffFIG} are easy to understand from the basic physics of brane-induced gravity. The scale-indepedent suppression of $G_{\rm eff}$ arises from the first term within braces in Eq.~(\ref{PoissonQSlin}), which is clearly always larger than unity, while the scale dependence arises through the second term that is always positive, enhancing the previous effect at large scales. 

The physical origin of these terms is very simple. They arise because the gravitational effect of sources at the brane is not only to create 4D intrinsic curvature (encoded by the 4D Ricci scalar), but also to create extrinsic curvature, which at small scales is dominated by the brane-bending mode $C_0$ (first term within braces in Eq.~\ref{PoissonQSlin}) and at larger scales by normal derivatives (second contribution within braces in Eq.~\ref{PoissonQSlin}), see Eq.~(\ref{dK}).

The precise fraction that goes into the usual 4D Ricci scalar (available to accelerate dark matter particles and thus grow structure) versus extrinsic curvature is controled by the Gauss-Codazzi equation, Eq.~(\ref{Eq44}), that relates intrinsic to extrinsic curvature. Therefore, as long as the extrinsic curvature remains important, density perturbations will have suppressed clustering because part of their ability to induce intrinsic curvature (the only type available in GR) is being used to create extrinsic curvature of the brane. We will see in the next section that nonlinear effects suppress the extrinsic curvature at small scales leading to the recovery of GR.

\begin{figure}[t!]
\begin{center}
\includegraphics[width=0.5\textwidth]{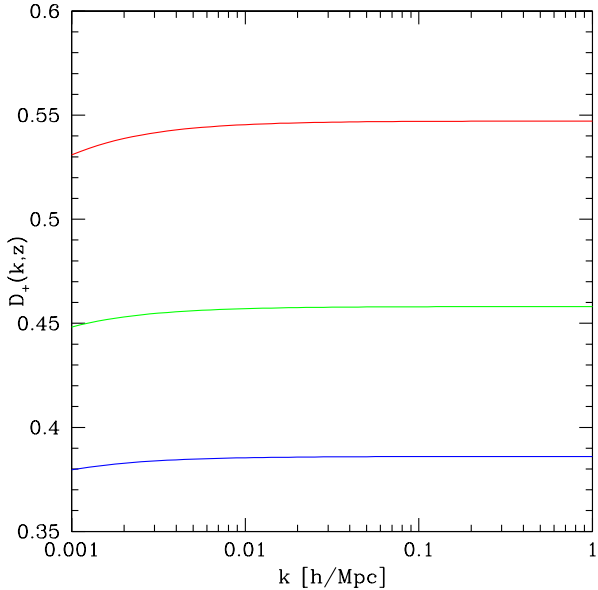}
\caption{The growth factor in the quasistatic approximation as a function of $k$ for different redshifts $z=0,0.5,1$ (from top to bottom).}
\label{DplusLinFIG}
\end{center}
\end{figure}

From the modified Poisson equation, Eq.~(\ref{PoissonQSlin}), and the conservation equations, Eqs.~(\ref{DivT0}-\ref{DivTi}), at subhorizon scales we have the linear density perturbation growth equation,

\beq
{d^2\d \over d\tau^2}+{\cal H} {d\d \over d\tau}=\nabla^2\phi_0=4\pi G_{\rm eff}(k,\eta) a^2 \bar{\rho}\, \delta
\label{linpt}
\eeq
where $\tau$ is conformal time ($ad\tau=dt$) and ${\cal H}=aH$ is the conformal expansion rate. Changing from conformal time to our time variable, $d\eta={\cal H}(q-1)\eta\, d\tau$, we can find the growth factor as a function of time and scale. The results are shown in Fig.~\ref{DplusLinFIG}. The overall behavior of the growth factor is expected given the properties of $G_{\rm eff}$ shown in Fig.~\ref{GeffFIG}, the scale dependence is much smaller due to the fact that the growth results from the integrated history of the strength of gravity and for most of the evolution the scale dependence in $G_{\rm eff}$ is small. Our results are in good agreement with those in~\cite{2007PhRvD..75f4002S,2008PhRvD..77h3512C} for the scale dependence at large scales, suggesting that most of the effect is due to the normal derivatives in the quasistatic approximation.

\section{Beyond Linear Theory}
\label{NL}

\subsection{Nonlinearities}

We now discuss how the linear results derived in the previous sections change when nonlinear effects are included. The first step is to identify the sources of new nonlinearities not present in GR. Looking at the action in ADM form, Eq.~(\ref{Sgrav}), the new contributions are in the second term proportional to the lapse function $N$.  We see right away that  new relevant nonlinearities appear through the 44 field equation, Eq.~(\ref{Eq44}), where the extrinsic curvature is  $\delta K \propto \nabla^2C$, see Eq.~(\ref{dK}). Since  $\nabla^2C_0 \propto \delta$ in linear theory, clearly the quadratic terms in Eq.~(\ref{Eq44}) will become important as $\delta$ becomes of order unity. Neglecting subleading terms the nonlinear version of Eq.~(\ref{Eq44lin}) becomes
\beq
(2\alpha'+\beta')\, \nabla^2 C +\frac{{\rm e}^{-2\alpha}}{2} [ (\nabla^2C)^2-(\nabla_{ij}C)^2] = \nabla^2\Psi,
\label{Eq44BNLlocal}
\eeq
which agrees with the equation written down in~\cite{2007PhRvD..75h4040K} for the brane-bending mode; see Eq.~(\ref{44NL0}) below for a version of this equation at the brane including normal derivatives. Therefore the dynamics of the brane-bending mode is nonlinear~\cite{2002PhRvD..65d4026D,2003AnPhy.305...96A,2003JHEP...09..029L,2004JHEP...06..059N,2004PhRvD..69l4015L,2004PhRvD..69b4001T,2007PhRvD..75h4040K}; this is crucial as it determines the relationship between the Newtonian potential $\phi_0$ and the spatial curvature potential $\psi_0$, through Eq.~(\ref{CBC}), which it is easy to check {\em does not} receive significant corrections even in the nonlinear regime. These two facts together determine how the theory reduces to GR at small scales, we'll explain this in more detail in the next sections.
 
There are other, similar, nonlinear corrections that affect the bulk solutions derived in the previous section. The main reason is that while $\Phi$ obeys an approximately decoupled equation from $C$ at the linear level due to its properties  under gauge transformations involving the extra dimension, when nonlinear effects are included this result no longer holds, as fields of different symmetry are coupled. Thus there is a coupling between fields being approximately constant in the bulk ($\Psi,C$) and those that decay strongly ($\Phi,f$). For example, Eq.~(\ref{ijbulk}) becomes ($k\neq i,j$)
\begin{widetext}
\beq
\nabla_{ij} [3 C' + 2(2\alpha'+\beta')\, C - 2\Psi+\Phi ]  + 3 {\rm e}^{-2\alpha}\ [ \nabla_{ij}C\, \nabla_{kk}C-\nabla_{ik}C\, \nabla_{jk}C]
= 0,
\label{ijbulkNL}
\eeq
\end{widetext}
therefore there are corrections of order $(k/aH)^2 HC$ to Eq.~(\ref{ijbulk}) and Eq.~(\ref{CPhi}), with the consequence that the details of how $C$ behaves in the bulk will be modified somewhat. In second-order perturbation theory, is easy to see that a new contribution develops that makes $C$ decay slowly into the bulk as ${\rm e}^{-\alpha}$, this is expected to kick in at high $k$. 

Similarly, the bulk equation for $\Phi$, Eq.~(\ref{PhiBulkQS}), develops in second order a source term going as $\sim {\rm e}^{-2\alpha}\, (\nabla^2C)^2$, which again becomes important at high-$k$, meaning that $\Phi$ will not decay as strongly into the bulk as given by linear theory, Eq.~(\ref{QSsol}). This means that the normal derivatives at high-$k$ will not be as large as given by Eq.~(\ref{normder}), which in turn implies that the nonlocal terms in the equations of motion will go away at small scales faster than expected based on linear theory. Summarizing, these nonlinear effects will make small nonlocal terms even more subdominant in the nonlinear regime. Therefore, we conclude it is safe to neglect these nonlinearities in the bulk equations.

Finally, there are other possible nonlinear terms that involve the interaction of $C$ with the other metric perturbations. Since it is precisely the brane-bending mode $C$ that is a ghost (at least in the de Sitter limit), one concern is that interaction of $C$ with normal fields can develop a catastrophic instability of cosmological solutions (see e.g.~\cite{2003PhRvD..68b3509C,2004PhRvD..70d3543C} for toy models and discussion of quantum effects). 

The most dangerous of such interaction terms that correct e.g. Eq.~(\ref{Eq44lin}) have the form $\Phi' \nabla^2C$ to quadratic order and $\Phi \Phi' \nabla^2C$ to cubic order. Given that $\Phi' $ does not grow with $k$ at high-$k$ due to the nonlinear effects discussed above and that its power spectrum decreases rapidly with $k$ ($P_\Phi \propto k^{-5}$), these terms do not give significant corrections on any cosmological scales, since $\Phi\ll1$. In regions where the weak field approximation breaks down ($\Phi \simeq 1$), e.g. near a black hole, the situation can of course be very different. Although in such high density regions, one may have difficulties exciting $C$ due to the Vainshtein suppression, and since $C$ does not obey a superposition principle, one may not add ``free" solutions to cosmological ones, thus any analysis must be done self-consistently. See Fig.~7 in paper~II~\cite{LSSbrane2} for a self-consistent solution showing how a cosmological background is enough to suppress $C$ at very large scales compared to astrophysical objects.

\subsection{Vainshtein Mechanism for Cosmological Perturbations}

Having identified the main new sources of nonlinearities, we are ready to discuss their effect. Before we do that in detail, it is useful to give a simple picture in geometrical terms of the role of nonlinearities, which lead to the recovery of GR at small scales, by the Vainshtein mechanism~\cite{1972PhLB...39..393V}. For detailed discussion of this in the context of the Schwarzschild solution see
~\cite{2002PhRvD..65d4026D,2002PhRvD..66d3509L,2002PhLB..534..209P,2004PhRvD..69b4001T,2005NewA...10..311G,2005PhRvD..72h4024G,2009arXiv0901.0393B}, here we are rather interested in how the mechanism works in a cosmological setting, for previous discussion along these lines see~\cite{2004PhRvD..69l4015L,2007PhRvD..75h4040K}.

We start from the field equations {\em at the brane}, which for 4D components can be written as
\beq
(R_{\mu\nu}-{1\over 2}g_{\mu\nu} R)-{1\over r_c}(K_{\mu\nu}-g_{\mu\nu}K) = 8\pi G T_{\mu\nu},
\label{RKT}
\eeq
where the first term in parenthesis is the usual Einstein tensor characterizing intrinsic curvature, and the second term denotes the contribution coming from extrinsic curvature of the brane embedded in the bulk. The relationship between intrinsic and extrinsic curvatures is given by Eq.~(\ref{Eq44}), and valid everywhere in the bulk (including arbitrarily close to the brane, which is what we are interested in this section).  Roughly speaking, we can interpret the self-accelerated solution by saying that for this branch the extrinsic curvature is of the same sign as intrinsic curvature, leading to the Friedmann equation Eq.~(\ref{FeqDGP}) and a self-accelerated universe where intrinsic curvature ($\sim H^2$) is balanced against extrinsic curvature of the brane ($\sim H$) when density of matter has dropped sufficiently. The threshold extrinsic curvature at which this transition happens is given by the inverse curvature radius $r_c^{-1}$. Therefore, the universe accelerates when extrinsic curvature starts to play a role.

We can understand the behavior of perturbations and the recovery of GR at small scales in similar, geometric terms. For this purpose let's rewrite Eq.~(\ref{Eq44}) in the following symbolic form (hereafter we also drop numerical factors of order unity)

\beq
R=\bar{R}+\delta R \sim K^2 = (\bar{K}+\delta K)^2
\label{Eq44simple}
\eeq
where quantities with bars refer to background values. Since $\bar{R}=\bar{K}^2 \sim H^2$, for perturbations we have

\beq
\delta R \sim  \bar{K} \delta K + (\delta K)^2,
\label{deltaR}
\eeq
with $\delta R \sim\nabla^2\Psi/a^2$ and $\delta K \sim \nabla^2C/a^2$, assuming we are at scales well within the Hubble radius. We can then write

\beq
{\delta R \over \bar{R}} = {\delta K \over \bar{K}} + \Big({\delta K \over \bar{K}}\Big)^2
\label{RKfluct}
\eeq
where 
\beq
{\delta R\over \bar{R}} \sim \Big({k\over aH}\Big)^2\Psi \sim \delta,
\eeq
\beq
{\delta K\over \bar{K}}\sim \Big({k\over aH}\Big)\Big({k\over a}\Big)\, C
\label{dKC}
\eeq

Then at large scales, the solution of Eq.~(\ref{RKfluct}) is
\beq
{\delta R \over \bar{R}} \sim {\delta K \over \bar{K}} \sim \delta \ll 1
\eeq
and fluctuations in intrinsic and extrinsic curvature are comparable, thus gravity is modified by the contribution of extrinsic curvature fluctuations encoded in the brane bending mode $C$. The same is true at scales approaching the Hubble radius when normal derivatives dominate the contribution to the extrinsic curvature, leading to 5D behavior.

When $\delta$ becomes of order unity, however, the second term in Eq.~(\ref{RKfluct}) becomes important. Therefore, at small scales we have instead, 
\beq
{\delta R \over \bar{R}} \sim \Big({\delta K \over \bar{K}}\Big)^2 \sim \delta \gg 1
\eeq
and fluctuations in intrinsic curvature ($\sim \delta$) are much larger than in extrinsic curvature ($\sim \delta^{1/2}$). Thus fluctuations in extrinsic curvature of the brane are suppressed at small scales and GR is restored.

Note that although fluctuations in extrinsic curvature become large in the nonlinear regime, there is no metric perturbation that is of order unity~\cite{2002PhLB..534..209P}. One can see that from Eq.~(\ref{dKC}), while $\delta K / \bar{K}$ may be large, $g_{i4} = \nabla_i C$ is always small, since $k/aH$ is very large in the nonlinear regime. We will explicitly check this with N-body simulations in paper II.

We can put the condition for transition to GR, $\delta K / \bar{K} \simeq 1$ in a more familiar form by considering spherical isolated perturbations, for which $\phi \sim r_g/r$ where $r_g$ is the gravitational radius corresponding to the mass enclosed by the perturbation~\cite{2004PhRvD..69l4015L}. Since $C\sim r_c \phi$, using Eq.~(\ref{dKC}) we can translate the requirement $\delta K / \bar{K} \simeq 1$ to find the scale $r_*$ away from the perturbation where the transition to GR takes place

\beq
{1\over H}\,{1\over r_*^2}\, r_c \,{r_g\over r_*} \sim 1,
\eeq
which gives the Vainshtein scale $r_*^3 \sim r_c H^{-1} r_g \sim r_c^2 r_g$ familiar from the Schwarzschild solution~\cite{2002PhRvD..65d4026D,2002PhRvD..66d3509L,2002PhLB..534..209P,2004PhRvD..69b4001T,2005NewA...10..311G,2005PhRvD..72h4024G,2009arXiv0901.0393B}. The difference here is that in the cosmological context $r_g$ is  the gravitational radius corresponding to the mass enclosed by the perturbation, not the total mass including the background, whose gravitational effects are already taken into account in the evolution of the Hubble constant~\cite{2004PhRvD..69l4015L}. Indeed, although at small scales the relation between metric and stress-energy fluctuations approaches GR, perturbations still evolve in a background that has a modified expansion history. The difference between total mass and perturbation mass becomes totally negligible for astrophysical objects within the range of their influence as the mass contributed by the background is completely negligible. For cosmological perturbations, the opposite is the case for long wavelength perturbations and thus while counting total mass one would expect no corrections to GR below scales $H^{-1}$~\cite{2006NJPh....8..326D,2008CQGra..25o4008G} (and nonlinearities at that scale), properly taking into account the background one expects that for long-wavelength perturbations $\lambda > r_*$ and thus modifications to their growth. 

To obtain the precise value for $r_*$ in a given cosmology one has to solve the nonlinear equation Eq.~(\ref{RKfluct}) (which is meant to  represent Eq.~\ref{Eq44}). The answer for the Vainshtein scale $r_*$ can only be given in a statistical sense, that is, the solution of that equation will depend on the spectrum of perturbations and cosmological parameters. In the context of the Schwarzschild solution, one can think of it as having to estimate $r_*$ from a collection of objects (how many and of what masses depending on the spectrum of perturbations) in a situation where the superposition principle does not apply (hence it is not possible to use the known answer for a single object). How different objects interact is controlled by the nonlinearities. We will tackle the calculation of the Vainshtein scale using perturbation theory below, and N-body simulations in paper~II~\cite{LSSbrane2}. See~\cite{2009arXiv0905.2966H} for a recent discussion of the impact of Vainshtein's mechanism for macroscopic violations of the equivalence principle.

\subsection{The Effective Poisson Equation}

Let us now discuss in detail the behavior of perturbations including nonlinearities. We can incorporate the nonlinear dynamics of the brane-bending mode into the Poisson equation and thus find an effective Poisson equation that relates the Newtonian potential to the density perturbation. Equation~(\ref{Eq44}) at the brane becomes (generalizing Eq.~\ref{44lin0}),

\begin{widetext}
\beq
H(q+2)\, \nabla^2  C_0  +
Hqa\sqrt{-\nabla^2} \, \Phi_0 - \nabla^2 \Psi_0 =
\frac{1}{2a^2} \Big[ (\nabla_{ij}C_0)^2 - (\nabla^2C_0)^2\Big]
\label{44NL0}
\eeq
which we can write in Fourier space as,

\beq
-{\eta(q+2)+1\over r_c}\, k^2 C_0 + Hqak\, \Phi_0 + k^2\, \phi_0 = {1\over 2a^2} 
\int \Big[(\k_1 \cdot \k_2)^2-k_1^2 k_2^2 \Big]\, C_0(\k_1)C_0(\k_2)\,  [\dD] \, d^3k_1d^3k_2,
\label{44NL0F}
\eeq
where $[\dD] \equiv \delta_{\rm D}(\k-\k_1-\k_2)$. This, together with the Poisson equation, Eq.~(\ref{PoissonQSprelim}), which in view of Eq.~(\ref{CBC}) we can rewrite in terms of $\phi_0$ as
\beq
-k^2\phi_0-\frac{1}{2r_c}k^2C_0 -\frac{ak}{2r_c}\Phi_0 = 4\pi Ga^2\bar{\rho}\, \delta,
\label{PoissonQSnew}
\eeq
determines how gravity responds to density perturbations. After expressing all metric variables interms of $\phi_0$ and $C_0$, using that $\Phi_0=2\phi_0+[\eta(q-1)+1]C_0/r_c$, we can use
Eq.~(\ref{PoissonQSnew}) to eliminate the brane-bending mode $C_0$ in terms of $\phi_0$ and $\d$ and thus Eq.~(\ref{44NL0F}) results in an effective nonlinear Poisson equation that relates the gravitational potential $\phi_0$ to the density contrast $\d$,

\beq
-\Big(\Lambda\d+k^2 \phi_0 \Big)= \Big({\d G \over G}\Big)\Bigg\{
 \Lambda\, \d -  2\Big({ r_c \over a} \Big)^2 \int
\Big[1-(\tk_1 \cdot \tk_2)^2\Big]
 \Big(\Lambda\d + k_1^2\phi_0\Big) \Big(\Lambda\d + k_2^2\phi_0\Big)
\, [\dD]\, d^3k_1 d^3k_2 \Bigg\},
\label{NLPoisson}
\eeq

\end{widetext}
where
\beq
\d G\equiv G_{\rm eff}-G, \ \ \ \ \ \Lambda \equiv 4\pi G a^2 \bar{\rho} ,
\eeq
and we have used the fact that when the nonlinear term becomes important the normal derivative terms can be neglected, i.e. $a/(kr_c) \ll 1$ at small scales. 

It's easy to see that this equation simplifies considerably in the spherical approximation. Indeed, in that case the kernel $[1-(\tk_1 \cdot \tk_2)^2]= 2/3$ and Fourier transforming back to real space, Eq.~(\ref{NLPoisson}) becomes a local quadratic equation for $(\nabla^2\phi_0-\Lambda \delta)$ with source $\delta$, whose solution is

\beq
\nabla^2\phi_0 = \Lambda \delta + {3 \Lambda\over 4 g}\Big({\delta G \over G}\Big)\, \Big[ \sqrt{1+{8\over 3}g\, \delta}-1\Big],
\label{PoissonSph}
\eeq
where $g$ is a dimensionless nonlinearity parameter given by 
\beq
g \equiv  2 \Big({\d G \over G}\Big)^2 \Big({r_c \over a} \Big)^2 \Lambda  
 =  \frac{ (2\eta-1)^2\, \eta(\eta-1)}{3(2\eta^2-2\eta+1)^2},
\label{g}
\eeq
and {\em decreases} as time goes on (due to the dilution in matter density) from $g=1/3$ at early times to $g \approx 0.1$ at $z=0$. As a result, the transition to standard gravity happens at relatively large density contrasts $\delta \ga 10$. One can also interpret the spherical modified Poisson equation, by defining a $\delta$-dependent Newton's constant by 
\beq
{G_\delta \over G} \equiv {1\over \Lambda} {\partial \nabla^2\phi_0 \over \partial \delta}= 1 + {(\delta G/G)\over \sqrt{1+{8\over 3}g\, \delta}},
\label{Gdelta}
\eeq
which also emphasizes that for voids, $\delta<0$, $G_\delta$ is further suppressed (recall $\delta G/G<0$ in our case) compared to $\delta>0$, that's another way of saying that new non-Gaussianities will be induced. 

Equation~(\ref{PoissonSph}) is the same as the modified Poisson equation given by Eq.~(3.1) in~\cite{2004PhRvD..69l4015L}, after using that their variables $\epsilon$ and $\beta$ can be written in our notation as $\epsilon=(8/3)g\,\delta$ and $\beta=3\, (\delta G/G)$. Note, however, that our treatment includes the scale dependence of $\delta G/G$ induced by the normal derivatives to the brane at the linear level, neglected in~\cite{2004PhRvD..69l4015L}.

In paper~II we also perform numerical simulations under the spherical approximation (used in~\cite{2009arXiv0903.1292K} to simulate DGP and degravitation theories) and compare them to the full solution to check their validity. We find that the spherical approximation is accurate to about 10\% in the density power spectrum.

Equation~(\ref{NLPoisson}) gives, beyond the spherical approximation, a {\em nonlocal} and nonlinear relationship between $\nabla^2\phi_0$ and the density perturbations $\delta$. We will explore two consequences of this equation. First, we discuss how to calculate the relationship between $\phi_0$ and $\delta$, i.e. the propagator for Eq.~(\ref{NLPoisson}), using an approximate resummation technique that captures the non-local corrections beyond the spherical approximation. We will use this to calculate the nonlinear power spectrum and evaluate the transition scale to GR. Finally, we use perturbation theory in Eq.~(\ref{NLPoisson}) to calculate the newly induced non-Gaussianities in the bispectrum.

\subsection{The Modified Poisson Propagator}

The right hand side of Eq.~(\ref{NLPoisson}) describes the corrections to the standard Poisson equation. At large scales, density perturbations are small and the first, linear, term dominates over the second, nonlinear, contribution. This is what leads to the modified growth of perturbations described by an effective scale and time dependent gravitational constant. As small scales are approached, however, the nonlinear term becomes important, and as we discussed above it makes the right hand side of Eq.~(\ref{NLPoisson}) subdominant compared to the left hand side, recovering GR at small scales.

To follow the nonlinear growth of structure we are interested in how $\nabla^2\phi_0$ responds to density fluctuations $\delta$. Since we are interested in calculating statistical averages (e.g. the power spectrum) we can describe the response  in Fourier space as 
\beqa
\phi_0 &=& \Big<{{\cal D}\phi_0(\k) \over {\cal D} \delta(\k')}\Big> \, \delta(\k')
 + {1\over 2!}\Big<{{\cal D}^2 \phi_0(\k) \over {\cal D} \delta_1 {\cal D} \delta_2}\Big> \, \delta_1\delta_2  \nonumber \\ & + & 
 {1\over 3!}\Big<{{\cal D}^3 \phi_0(\k) \over {\cal D} \delta_1 {\cal D} \delta_2  {\cal D}\delta_3}\Big> (\delta_1\delta_2\delta_3- \lexp \delta_1 \delta_2 \rexp \delta_3 - {\rm cyc.})
 \nonumber \\  &  + & \ldots
 \label{ResF}
\eeqa
where ${\cal D}$ denotes a functional derivative, integration over repeated Fourier arguments is understood, and $\delta_i\equiv \delta(\k_i)$. In the language of renormalized perturbation theory~\cite{2006PhRvD..73f3519C,2008PhRvD..78j3521B} the coefficient of the linear term can be regarded as the two-point propagator for the modified Poisson equation, 
\beq
\Gamma_\phi(k)\, \delta_{\rm D}(\k-\k') \equiv \Big<{{\cal D}\phi_0(\k) \over {\cal D} \delta(\k')}\Big>,
\label{Gamma}
\eeq
the coefficient of the quadratic term as its three-point propagator
\beq
\Gamma_\phi^{(2)}(\k_1,\k_2)\, \delta_{\rm D}(\k-\k_{12}) \equiv {1\over 2!}\Big<{{\cal D}^2 \phi_0(\k) \over {\cal D} \delta(\k_1) {\cal D} \delta(\k_2)}\Big>,
\label{Gamma2}
\eeq
and so on. The two-point propagator $\Gamma_\phi$ describes the renormalized linear response, with asymptotics
\beqa
\Gamma_\phi(kr_*\ll 1) & \simeq & -{\Lambda \over k^2}\, {G_{\rm eff} \over G}, \\
\Gamma_\phi(kr_*\gg 1) & \simeq & -{\Lambda \over k^2},
\label{GammaAsymp}
\eeqa
where $r_*$ is the Vainshtein scale. Recall that at large scales the effective Newton's constant $G_{\rm eff}$ depends on $k$, with nonlocal terms giving $\Gamma_\phi(k\rightarrow 0) \propto r_c k^{-1}$, in precise correspondence with the graviton propagator. On the other hand, at small scales where extrinsic curvature fluctuations are subdominant compared to density (intrinsic curvature) fluctuations $\Gamma_\phi$ returns to its GR (unmodified Poisson) value. The transition happens through nonlinear effects in the dynamics of the brane bending mode, which is the nontrivial part of evaluating the expectation value in Eq.~(\ref{Gamma}).

\begin{figure}[t!]
\begin{center}
\includegraphics[width=0.5\textwidth]{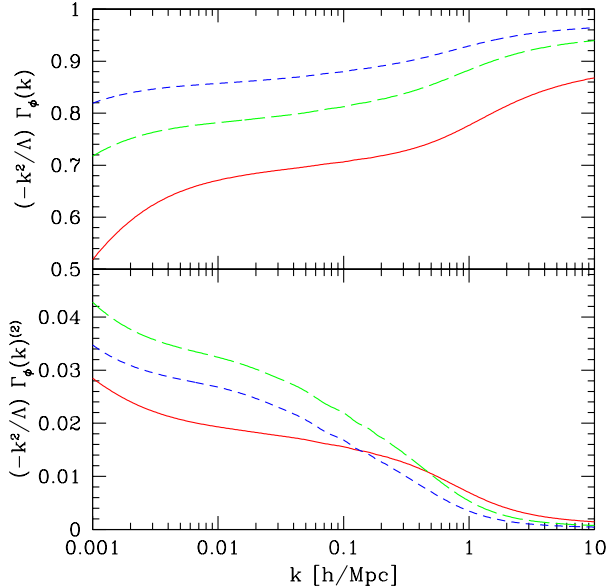}
\caption{{\em Top panel:} The two-point propagator for the modified Poisson equation in the naive spherical approximation, Eq.~(\protect\ref{GaSph}), for different redshifts corresponding to $z=0$ (solid), $z=0.75$ (long dashed), and $z=1.5$ (short dashed). {\em Bottom panel:} Same for the three-point propagator, Eq.~(\protect\ref{Ga2Sph}). Note that the time dependence is not monotonic in this case. }
\label{Gamma_Gamma2}
\end{center}
\end{figure}

The three-point propagator $\Gamma^{(2)}_\phi$ is important at large scales where it gives rise to additional non-Gaussianity. However, as we shall see below it is a rather small correction (few percent) at large scales, whereas at small scales we know from the discussion on the Vainshtein mechanism above that it must be highly suppressed since corrections to linear response at small scales go roughly as extrinsic curvature fluctuations which scale as $\sim \delta^{1/2}$, which from Eq.~(\ref{ResF}) suggest $\Gamma^{(2)}_\phi \sim\Delta^{-3/2}$, where $\Delta^2(k)\equiv 4\pi k^3 P(k)$ is the amplitude of density perturbations. Indeed, it is easy to see this in the spherically symmetric case. From Eq.~(\ref{PoissonSph}) we can roughly estimate
\beq
\Gamma_\phi (k) \approx -{\Lambda \over k^2}\ \Bigg[ 1+{(\delta G / G) \over \sqrt{1+{8\over 3}g \Delta(k)}}\Bigg],
\label{GaSph}
\eeq
where we naively replaced $\delta \rightarrow \Delta(k)$ when taking expectation values in Eq.~(\ref{Gdelta}). This is of course not a valid calculation of the spherical propagator (the dependence on the power spectrum is simply a guess, and higher-order statistics than the power spectrum should enter into the correct spherical answer), but it is enough for estimation purposes. To make this distinction clear we refer to this calculation as the naive spherical approximation. Similarly, we can estimate the three-point propagator as
\beq
\Gamma_\phi^{(2)}(k) \approx  {\Lambda \over k^2}\times{2\over 3}g\, \Big({\delta G\over G}\Big) \Big[1+{8\over 3}g\, \Delta(k) \Big]^{-3/2}.
\label{Ga2Sph}
\eeq
Thus we see that when $\Delta \gg 1$, $\Gamma^{(2)}_\phi \sim\Delta^{-3/2}$ and that $\Gamma^{(2)}_\phi \ll \Gamma_\phi$ at all scales. The overall scale of $\Gamma^{(2)}_\phi$ compared to $\Gamma_\phi$ is set by the nonlinear coupling $g$, which is always much smaller than unity (see Eq.~\ref{g}). Higher-order propagators will be suppressed by correspondingly more powers of $g$, making the description of the modified Poisson equation by the two-point propagator alone a very good approximation.

Figure~\ref{Gamma_Gamma2} shows these results as a function of $k$ for  redshifts $z=0,0.75,1$, where for the power spectrum we have used the HaloFit fitting formula~\cite{2003MNRAS.341.1311S} for standard gravity corresponding to a linear normalization given by the DGP expansion history. We will compare these simple results with the calculation including effects beyond the spherical approximation that follows below.

All these arguments suggest that to obtain an effective resummation of the two-point propagator $\Gamma_\phi$ is very reasonable to use linear response in Eq.~(\ref{NLPoisson}). For convenience in writing the equations that follow below we write the propagator in terms of a new function $M(k)$, 
\beq 
\Lambda\d(\k)+k^2 \phi_0(\k) \equiv - \Lambda\, \Big({\d G \over G}\Big)(k)\,  \d(\k) \, M(k),
\label{meanF}
\eeq
thus
\beq
\Gamma_\phi(k) = - {\Lambda \over k^2} \ \Big[1+{\delta G \over G}\, M(k) \Big],
\label{GammaM}
\eeq
where $M$ describes {\em on average} the Vainshtein nonlinear dynamics, with asymptotics
\beq
M(k\rightarrow0) \rightarrow 1,\ \ \ \ \ M(k\rightarrow\infty)\rightarrow0.
\label{Tasymp}
\eeq
We must note that the resummation is essential to recover the large-$k$ limit, and that $M(k)$ in Eq.~(\ref{meanF}) denotes effective (in mean field sense) corrections to the Poisson equation, which is what enters when calculating correlation functions; i.e. it is not meant to describe a particular realization in the ensemble (for that it would have to be a random field itself). We may regard the function $M(k)$ as a form-factor that describes the suppression of the brane-bending mode at small scales; or, in terms of the graviton propagator, it corresponds to including such momentum dependent form-factor in the second term on the second line of Eq.~(\ref{Solvehmunu}), corresponding to the helicity-zero mode, thus leading to the disappearance of the vDVZ discontinuity at high momentum.

The nonlinearity of the effective Poisson equation is encoded in the fact that $M$ depends on the spectrum and bispectrum of density perturbations, while the non-locality essentially comes from the fact that the bispectrum has a nontrivial dependence on triangle shape. Although we only explicitly used $k$ as an argument, $M$ also depends on time, as do all variables in Eqs.~(\ref{meanF}-\ref{GammaM}). In order to find an equation for $M$ we rewrite Eq.~(\ref{NLPoisson}) using Eq.~(\ref{meanF}),

\begin{widetext}

\beq
(M-1)\,\d(\k)= - g   \int
\Big[1-(\tk_1 \cdot \tk_2)^2\Big]\,
\d_1 M_1 \, \d_2  M_2\, [\dD]\, d^3k_1 d^3k_2,
\label{NLP1}
\eeq
and multiplying by $\d(\k')$, taking expectation values, and Fourier transforming back to real space, we find

\beq
\xi(r)-\chi(r) = g  \int
\Big[1-(\tk_1 \cdot \tk_2)^2\Big] \, B(\k_1,\k_2)\, M_1  M_2\, {\rm e}^{i\k_{12}\cdot\r} \, d^3k_1 d^3k_2,
\label{NLP2}
\eeq
where the correlation functions are,
\beq
\xi(r) = \int {\rm e}^{i\k\cdot\r}\, P(k)\, d^3k, \ \ \ \ \ \chi(r) = \int {\rm e}^{i\k\cdot\r}\, P(k) \,M(k)\, d^3k, 
\label{xis}
\eeq
and $P$ and $B$ are the power spectrum and bispectrum of density perturbations,
\beq
\lexp \d(\k) \d(\k') \rexp= P(k)\, \dD(\k+\k') ,\ \ \ \ \ \lexp \d(\k_1)\d(\k_2)\d(\k_3) \rexp= B(\k_1,\k_2) \, \dD(\k_1+\k_2+\k_3).
\label{PB}
\eeq

As mentioned above, the non-Gaussian corrections induced by the nonlinearities in the modified gravity sector are small, since they are suppressed by powers of $g$ (see Section~\ref{BispSec} below for an explicit calculation of the bispectrum). Therefore, for the purpose of calculating $M(k)$ we shall use the standard gravity bispectrum. Since the Vainshtein scale is in the nonlinear regime~\cite{2004PhRvD..69l4015L}, we need a description of the bispectrum at nonlinear scales. Thus we use the bispectrum fitting formula obtained in~\cite{2001MNRAS.325.1312S}, which reads
\beq
B(\k_1,\k_2)=2F_2(\k_1,\k_2)P_1P_2 +2F_2(\k_2,\k_3)P_2P_3 +2F_2(\k_3,\k_1)P_3P_1,
\label{Btree}
\eeq 
where $\k_3=-\k_{12}$ and the kernel $F_2$ is given by
\beq
F_2(\k_1,\k_2)={5\over 7}\, a(k_1)a(k_2)+{1\over 2}(\tk_1 \cdot \tk_2)\Big({k_1\over k_2}+{k_2\over k_1}\Big)\, b(k_1)b(k_2)+ {2\over 7}(\tk_1 \cdot \tk_2)^2 \, c(k_1)c(k_2),
\label{F2}
\eeq
where the functions $a$, $b$ and $c$ are given by~\cite{2001MNRAS.325.1312S}
\beq
a= \frac{ 1 + \sigma_8^{-0.2} \sqrt{0.7\,Q_3(n)}\, (q/4)^{n+3.5}}{1+(q/4)^{n+3.5}},\ \ \  \ \
b= \frac{1+0.4\, (n+3)\,q^{n+3}}{1+q^{n+3.5}},\ \ \ \ \ 
c= \frac{1+4.5/[1.5+ (n+3)^4] \,(2q)^{n+3}}{1+(2q)^{n+3.5}},
\eeq
where $q=k/k_{nl}$ with $k_{nl}$ the scale at which the dimensionless linear power is unity, $n$ is the effective spectral index at scale $k$, and  $Q_3(n)=(4-2^n)/(1+2^{n+1})$. At large scales, $a,b,c \to 1$ and tree-level perturbation theory is recovered, at small scales $b,c \to 0$ and a hierarchical 
bispectrum with saturation value $Q_3(n)$ follows~\cite{1999ApJ...520...35S}. Possible deviations from the hierarchical ansatz in the highly nonlinear regime will only impact $M(k)$ at very large wavenumbers, where it is nearly zero. 

In order to solve for $M(k)$ from Eq.~(\ref{NLP2}) it is convenient to define an effective amplitude $Q_{\rm eff}(r)$,
\beq
Q_{\rm eff}(r)\ [\chi(r)]^2 \equiv \int
\Big[1-(\tk_1 \cdot \tk_2)^2\Big] \, B(\k_1,\k_2)\, M_1  M_2\, {\rm e}^{i\k_{12}\cdot\r} \, d^3k_1 d^3k_2,
\label{Qeff}
\eeq
\end{widetext}
which converts Eq.~(\ref{NLP2}) into a local equation for $\chi(r)$,
\beq
\xi(r)-\chi(r) = g \, Q_{\rm eff}(r)\ [\chi(r)]^2,
\label{LocEqn}
\eeq
and the solution for $M$ is obtained from the resulting quadratic equation for $\chi(r)$ by inverse Fourier transform,
\beq
M(k)=  \int   {d^3r \over(2\pi)^3}\, {\rm e}^{-i\k\cdot\r} \ 
\frac{\sqrt{1+4\,g\,Q_{\rm eff}\, \xi(r)}-1}{2g\,Q_{\rm eff}\, P(k)}.
\label{Msol}
\eeq
It's easy to see that this $M(k)$ satisfies the asymptotics in Eq.~(\ref{Tasymp}) and represents an effective resummation of the Vainshtein mechanism. The Vainshtein scale for cosmological perturbation thus depends on the characteristic nonlinearity $g$ (see Eq.~\ref{g}), the bispectrum of perturbations (since the nonlinearities in the field equations are quadratic) weighted by the nonlocal kernel of second derivatives $1-(\tk_1 \cdot \tk_2)^2$ represented here by $Q_{\rm eff}$, and the power spectrum of perturbations.

To carry out the calculation of $M$, we then proceed as follows:

\begin{itemize}

\item[i)] We start assuming the naive spherical solution for the modified Poisson propagator (and thus $M$), Eq.~(\ref{GaSph}). 

\item[ii)] Given this $M$, we calculate $Q_{\rm eff}$ from Eq.~(\ref{Qeff}) at the desired redshift. For the power spectrum, we use the nonlinear power spectrum given by HaloFit~\cite{2003MNRAS.341.1311S} corresponding to a linear spectrum in standard gravity with the growth factor that follows the modified expansion history, let us call this GRH. For  example, for the $\Omega_m=0.2$ model we use, we assume $\sigma_8=0.9$ at $z=0$ for GR, then $\sigma_8=0.754$ for GRH, and $\sigma_8=0.696$ for linearized DGP. Since the  $Q_{\rm eff}$ ratio involves equal number of power spectra in numerator and denominator, the assumed power spectrum should not affect the answer by much. As we shall see, the GRH power spectrum never differs from the fully nonlinear DGP model by more than 30\%.

\item[iii)] Evaluate the new $M(k)$ from Eq.~(\ref{Msol}). 

\item[iv)] With this new solution for $M(k)$ we go back of step i) and iterate these steps until convergence is achieved.
\end{itemize}

\begin{figure}[t!]
\begin{center}
\includegraphics[width=0.5\textwidth]{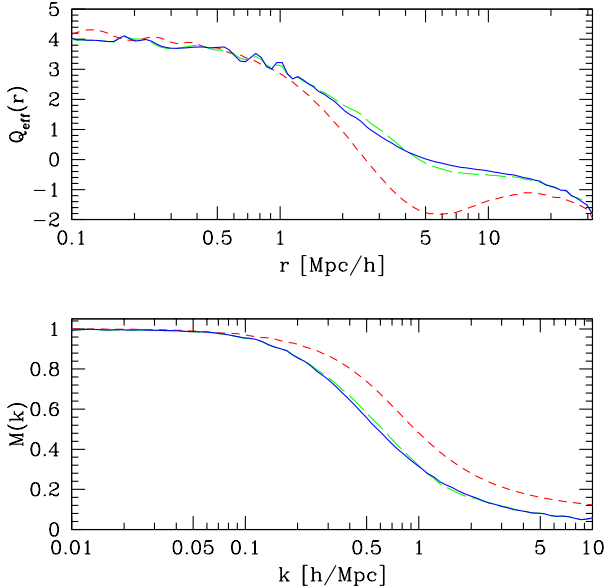}
\caption{Convergence of the iteration scheme to solve the integral equation for the modified Poisson propagator at $z=0$. Short dashed lines show the initial guess that corresponds to the naive spherical solution, long dashed lines shows the solutions after 3 iterations, and solid lines after 10 iterations. The top panel corresponds to $Q_{\rm eff}$, see Eqs.~(\protect\ref{Qeff}-\protect\ref{LocEqn}), while the bottom panel shows the function $M$ that defines the modified Poisson propagator, see Eqs.~(\protect\ref{GammaM}) and~(\protect\ref{Msol}).
}
\label{QeffMconv}
\end{center}
\end{figure}

Figure~\ref{QeffMconv} shows the results of this iterating procedure for $z=0$. Short dashed lines show the initial guess that corresponds to the naive spherical solution for $M$ (bottom panel) and that implied for $Q_{\rm eff}$ (top panel) after using Eq.~(\protect\ref{Qeff}), long dashed lines shows the solutions after 3 iterations, and solid lines after 10 iterations. We see that convergence is quite fast, typically a few iterations (at high redshift convergence is much faster). Some noise develops in $Q_{\rm eff}$ at small scales ($r \la 1 \Mpc$), but as we checked by fitting a smooth function to $Q_{\rm eff}$, this does not affect the behavior of $M(k)$. 
 
\begin{figure}[t!]
\begin{center}
\includegraphics[width=0.5\textwidth]{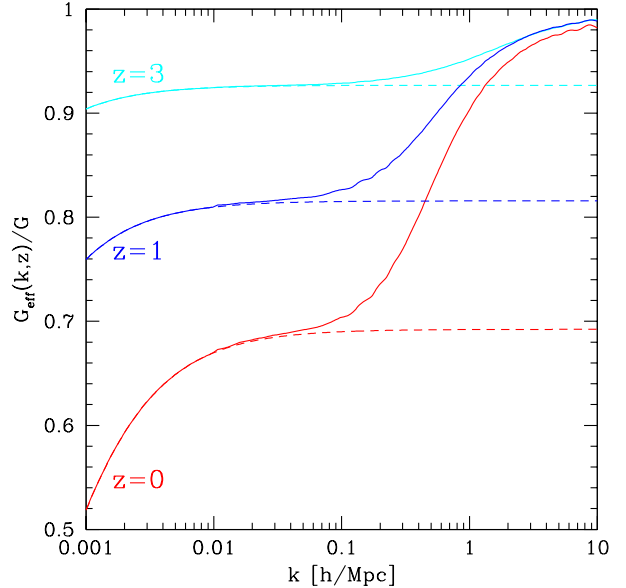}
\caption{Effective gravitational constant (solid line) as a function of scale at $z=0,1,3$ after resummation of nonlinearities in the brane-bending mode (Vainshtein mechanism). Dashed lines show the linear theory calculation, as in Fig.~\protect\ref{GeffFIG}.}
\label{GeffNL}
\end{center}
\end{figure}

Figure~\ref{GeffNL} shows the effective gravitational constant as a function of scale when the resummation of the modified Poisson propagator is included. We see that nonlinearities in the modified gravity sector drive $G_{\rm eff}/G$ to unity at small scales, as expected from the Vainshtein mechanism. We now discuss what this renormalization of the gravitational constant implies for the nonlinear power spectrum.

\section{Statistics}

\subsection{Nonlinear Equations of motion}

We now consider the equations of motion for density and velocity fields at the nonlinear level, and include the nonlinearities in the modified gravity sector, first in terms of the running gravitational constant (two-point modified Poisson propagator) and what this implies for the nonlinear power spectrum, and second by including the three-point propagator and what it implies for the bispectrum. Let us change to a time variable in terms of the scale factor $a$, to
\beq
x\equiv \ln a, 
\eeq
then we can rewrite conservation of stress-energy as

\beq
{\partial \delta \over \partial x} - \nabla \cdot \hat{\bf v} = \nabla (\delta \hat{\bf v}),
\label{ContNL}
\eeq
where we have written the velocity field as ${\bf v} \equiv - {\cal H} \hat{\bf v}$, and using $\theta \equiv \nabla \cdot \hat{\bf v}$,

\beq
{\partial \theta \over \partial x}+(1+q)\, \theta  - \bar{\nabla}^2 \phi_0 = \nabla \cdot [(\hat{\bf v}\cdot \nabla)\hat{\bf v}],
\label{EulerNL}
\eeq
where $ \bar{\nabla}$ is the dimensionless gradient,
\beq \bar{\nabla} \equiv \frac{\nabla}{aH}, \ \ \ \ \ \bar{k} \equiv \frac{k}{aH}  
\eeq
and we can write the modified Poisson equation as,


\beq
 \phi_0(\k) = \Gamma_\phi(k) \, \delta(\k) +  \int \Gamma^{(2)}_{\phi}(\k_1,\k_2)\, \d_1\d_2\, [\dD] d^3k_1d^3k_2.
\label{PoissonGamma}
\eeq

Equations~(\ref{ContNL}), (\ref{EulerNL}) and (\ref{PoissonGamma}) form the closed system of equations to be solved, and constitutes the starting point to compute nonlinear corrections to the power spectrum and bispectrum. We can write a second-order equation for the evolution of density perturbations (putting linear terms in the left hand side and quadratic nonlinearities in the right hand side),

\begin{figure}[t!]
\begin{center}
\includegraphics[width=0.5\textwidth]{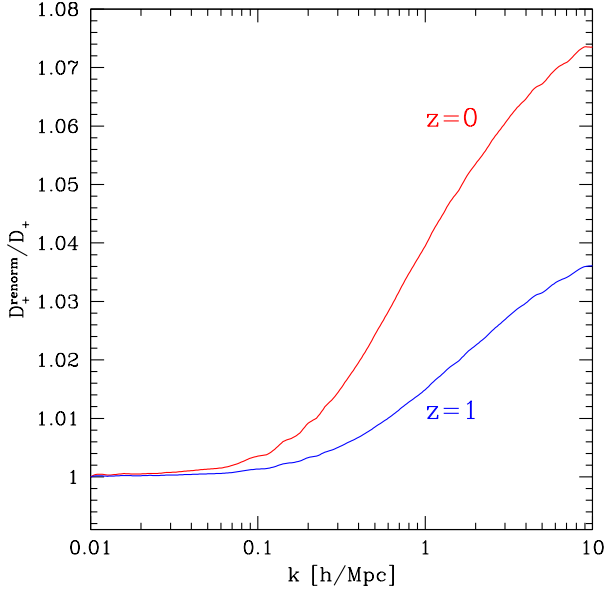}
\caption{Ratio of renormalized growth factor to standard linear growth factor as a function of scale for $z=0,1$. The enhancement towards small scales reflects the onset of the Vainshtein mechanism, by which gravity becomes GR.}
\label{DplusR}
\end{center}
\end{figure}

\begin{widetext}
\beq
{\partial^2 \delta \over \partial x^2}  +(1+q)\, {\partial \delta \over \partial x}  + \bar{k}^2\, \Gamma_\phi \, \delta = 
\Big[ (1+q)+{\partial \over \partial x} \Big] \nabla\cdot (\d \hat{\bf v} ) + \nabla \cdot [(\hat{\bf v}\cdot \nabla)\hat{\bf v}]
-  \bar{k}^2 \int \Gamma^{(2)}_{\phi}\, \d_1\d_2\, [\dD] \,d^3k_1d^3k_2
\label{EOMdelta}
\eeq
\end{widetext}
where we have slightly abused notation by writing some terms in real space (the standard gravity nonlinearities) and others in Fourier space (the modified gravity nonlinearities, which are resummed into the Poisson propagators $\Gamma_\phi$'s). This form of the equations of motion is useful to derive linear and second-order perturbative solutions.

At this point it is worth mentioning that had we carried out the full perturbative expansion of the DGP equations (see~\cite{2009arXiv0902.0618K}), we would end up with an infinite series of vertices corresponding to the perturbative solution of Eq.~(\ref{NLPoisson}) for $\phi_0$ as a function of $\delta$. These terms will have two effects when calculating correlation functions: they will renormalize the growth factor (essentially because one can dress any linear propagator inside a diagram by including such vertices: this is described by $\Gamma_\phi$), and they can change the mode-coupling properties (described by $\Gamma_\phi^{(2)}$ and higher-order ones). While the first effect leads to the Vainshtein mechanism through a dressing of the gravitational constant, the second is subdominant to the mode-coupling effects in GR, because the vertices are suppressed by powers of $g$ as opposed to those in GR which are always of order unity. Thus, while we will need the resummation of the two-point Poisson propagator to describe nonlinear scales, the three-point (and higher) propagators only affect non-Gaussianity weakly on large scales, and provide very small corrections to the power spectrum on intermediate scales.

\subsection{Power Spectrum}

From Eq.~(\ref{EOMdelta}) we see that one effect of nonlinearities in the modified gravity sector is to renormalize the linear growth factor (which can be obtained from this equation by using a $\Gamma_\phi$ with $M=1$ in Eq.~\protect\ref{GammaM}). As $M$ decays to zero at large $k$ (see Fig.~\ref{QeffMconv}) the effective gravitational constant runs to $G$  (see Fig.~\ref{GeffNL}) and gravity is enhanced, becoming GR in the large $k$ limit. This means that the renormalized growth factor will increase compared to the linear growth at small scales. Figure~\ref{DplusR} shows the result for the renormalized growth factor $D_+^{\rm renorm}$ obtained by solving Eq.~(\ref{EOMdelta})  with zero right hand side, in terms of the linear growth factor $D_+$ as a function of scale for $z=0,1$. We see the expected enhancement; the effect is larger at lower redshift where nonlinearities are stronger, as expected.  Because the two-point propagator obeys the right asymptotics at small scales, Eq.~(\ref{Tasymp}), the renormalized growth factor becomes the linear growth factor in GR with the same expansion history as DGP. That means, for example, that parametrizing the growth factor in terms of a growth index $\gamma$~\cite{2005PhRvD..72d3529L}, $\gamma$ will run with scale and time, becoming closer to the GR value at small scales and late times.

\begin{figure}[t!]
\begin{center}
\includegraphics[width=0.5\textwidth]{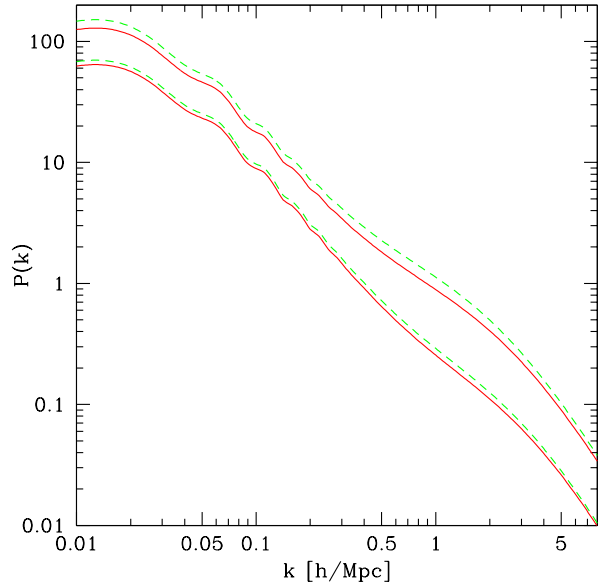}
\caption{The nonlinear density power spectrum at $z=0$ (top) and $z=1$ (bottom) for the DGP model (solid lines) and GR with the same expansion history (dashed lines).}
\label{PowerDGP}
\end{center}
\end{figure}

\begin{figure}[t!]
\begin{center}
\includegraphics[width=0.5\textwidth]{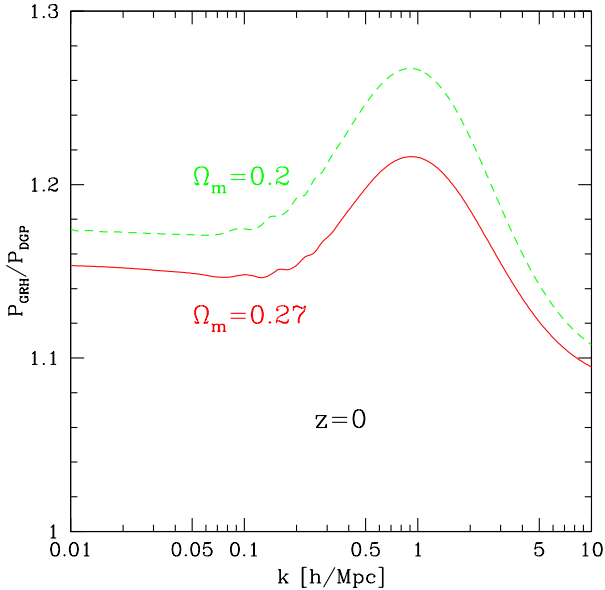}
\caption{The ratio of $P_{\rm GRH}$, the power spectrum for GR with a modified expansion history, to that in the DGP model for two values of $\Omega_m$, highlighting the dependence on the somewhat different Vainshtein scales.}
\label{OmDep}
\end{center}
\end{figure}

The right hand side of Eq.~(\ref{EOMdelta}) has the familiar form of standard gravity except for the last term containing the three-point propagator. As discussed above this is a small correction (suppressed by $g$) and decays very fast into the nonlinear regime. We shall quantify the size of this term in detail below when we discuss the bispectrum; those results indicate that the three-point propagator will induce extra mode coupling power at the 1-2\% level. Therefore, within the accuracy of our calculations here it is well justified to neglect this contribution when computing the nonlinear power spectrum.

Under this approximation, we are thus left with a rather remarkable situation. The whole effect of nonlinearities in the modified gravity sector is to renormalize the growth factor, or rather {\em renormalize the linear power spectrum}. To calculate the nonlinear power spectrum, we just have to evolve this renormalized linear spectrum by the standard nonlinearities we commonly deal with in GR. For this purpose, for simplicity we use the HaloFit~\cite{2003MNRAS.341.1311S} fitting formula that relates the linear spectrum to the nonlinear spectrum at given time. While this fitting formula was developed for CDM models with a linear evolution that is scale independent (and $G$ independent of scale and time), note that the time and scale dependence of $G$ in our case is rather small, its change is only appreciable over many Hubble times and orders of magnitude in spatial scales. The renormalized growth factor changes even more slowly than $G$, and an adiabatic adjustment of the linear power spectrum at each time and scale (as given by Fig.~\ref{DplusR}) should be a reasonable approximation. Note in addition that, clearly, our method correctly recovers large and small scales: at large scales nonlinearities are small and we recoverer full linear theory (including the transition to 5D behavior in the quasistatic approximation), at small scales where the Vainshtein mechanism makes gravity GR, our renormalized growth will bring the linear power spectrum to the amplitude it would have in GR with a modified expansion history. We shall compare our predictions to numerical simulations in detail in paper~II.

Figure~\ref{PowerDGP} shows the results for the nonlinear power spectrum at $z=0,1$. The dashed lines correspond to the GRH nonlinear spectrum, i.e. a model with the same expansion history (or Hubble constant) but with standard gravity~\footnote{In reality, we don't quite calculate $P_{\rm GRH}$ as we use the HaloFit $\Lambda$CDM fitting formula but with the linear normalization given by the modified Friedman equation. There is a correction that kicks in at intermediates scales due to the difference in linear decaying modes (which depend on expansion history), but that's a small effect~\cite{2006MNRAS.366..547M,2008JCAP...10..036P}.}. The difference in large-scale normalization ($\sigma_8=0.696$ versus $\sigma_8=0.754$ at $z=0$) reflects the difference in gravitational force law alone, and is stronger at lower redshift where the modified gravity has had more time to act. This difference gets enhanced at intermediate scales, but this is an expected outcome of the difference in normalizations, i.e. one-loop corrections would be stronger for $P_{\rm GRH}$ than for $P_{\rm DGP}$; the effect of the running $G$ in $P_{\rm DGP}$ is subdominant. At small scales the running of $G$ in $P_{\rm DGP}$ becomes important and $P_{\rm DGP}$ approaches $P_{\rm GRH}$. Some of these features are also seen in $f(R)$ models, see~\cite{2008PhRvD..78l3524O}.

Figure~\ref{OmDep} shows more clearly these effects by taking the ratio $P_{\rm GRH}/P_{\rm DGP}$ at $z=0$ for two different values of $\Omega_m$, the standard $\Omega_m=0.2$  that we used throughout this paper and $\Omega_m=0.27$ (which we use in paper~II to compare to simulations). Both models would have $\sigma_8=0.9$ at $z=0$ in $\Lambda$CDM. The lower $\Omega_m$ is, the larger $r_c$ is, and thus the larger the Vainshtein scale is, with stronger nonlinearities in the modified gravity sector (i.e. larger $g$). Therefore, although the difference between these two models is significant (up to 5\% for $k\la 1 \kvecMpc$), at smaller scales the low $\Omega_m$ model approaches $P_{\rm GRH}$ somewhat faster (the dashed line in Fig.~\ref{OmDep} approaches the solid line at large $k$).  Our results for the behavior of $P_{\rm GRH}/P_{\rm DGP}$ are qualitatively similar to those in~\cite{2009arXiv0902.0618K}, which use perturbation theory to estimate the quadratic correction to the Poisson equation and use it in a fitting formula proposed by~\cite{2007PhRvD..76j4043H} to interpolate between $P_{\rm DGP}$ for the linearized Poisson equation at large scales and $P_{\rm GRH}$ at small scales based on the amplitude of the dimensionless power $\Delta(k)=4\pi k^3 P(k)$.

\subsection{Bispectrum}
\label{BispSec}

\begin{figure}[t!]
\begin{center}
\includegraphics[width=0.5\textwidth]{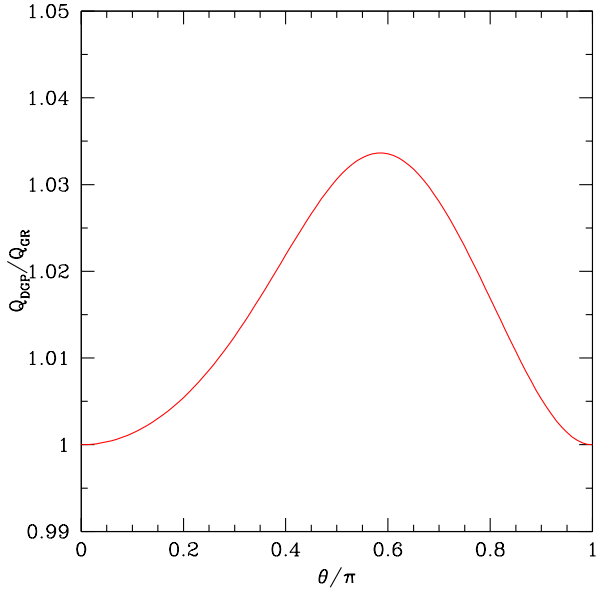}
\caption{The ratio of reduced bispectra $Q$ in brane-induced gravity to GR for triangles with $k_1=0.1 \kvecMpc$ and $k_2=2k_1$ as a function of angle $\theta$ between $\k_1$ and $\k_2$.}
\label{QDGPGR}
\end{center}
\end{figure}

Let us now discuss the new non-Gaussianities coming from the modified gravity sector, which are induced by the extrinsic curvature fluctuations. Since we are interested in large scales (but not as large so nonlocal terms become important), we can use the large-scale behavior of the Poisson propagators in Eq.~(\ref{EOMdelta}),
\begin{widetext}

\beq
\Gamma_\phi(k)\approx -{\Lambda \over k^2} {G_{\rm eff}\over G}, \ \ \ \ \ \ \ \ \ \ 
\Gamma_\phi^{(2)}(\k_1,\k_2) \approx g\, {\Lambda \over k^2} {\d G\over G}  \Big[1-(\tk_1 \cdot \tk_2)^2\Big]
\label{GammasLowk}
\eeq

To compute the large-scale bispectrum, we need the second-order solution to Eq.~(\ref{EOMdelta}). Since the new contributions do not add new multipoles (only changing the $\ell=0,2$ amplitudes), we follow the standard approach from PT (see e.g.~\cite{2004astro.ph..9224B}). We write the second-order solution as,

\beq
\delta^{(2)}(\k) = \int \Big[ {\cal D}_0 + {1\over 2} {\cal D}_1\, \Big({k_1\over k_2}+{k_2\over k_1}\Big) L_1(\tk_1 \cdot \tk_2) + 
{\cal D}_2\, L_2(\tk_1 \cdot \tk_2) \Big]\, \d_1\d_2\, [\dD] \, d^3k_1d^3k_2,
\label{delta2}
\eeq
where the ${\cal D}_i$ are arbitrary functions of time, and $L_\ell$ are Legendre polynomials. The terms inside the square brackets reflect the structure of the right-hand side of Eq.~(\ref{EOMdelta}) and correspond to the kernel $F_2$ in Eq.~(\ref{F2}) in the large-scale limit ($a=b=c=1$). From the constraint that $\lexp \delta^{(2)} \rexp =0$ it follows that ${\cal D}_1={\cal D}_0+{\cal D}_2$, and the two independent multipole amplitudes obey the following equations of motion (a dot denotes $d/dx$),

\beq
\ddot{{\cal D}}_0  +(1+q)\, \dot{{\cal D}}_0  - {3\over 2} \Big({\eta-1 \over \eta}\Big){G_{\rm eff}\over G} 
{\cal D}_0 = 
\Big[ (1+q)+{d \over d x} \Big] \Big( \dot{D_+}D_+\Big)+{1\over 3} (\dot{D_+})^2 
+   g{\d G\over G} \Big({\eta-1 \over \eta}\Big)D_+^2,
\label{EOMD0}
\eeq
\beq
\ddot{{\cal D}}_2  +(1+q)\, \dot{{\cal D}}_2  - {3\over 2} \Big({\eta-1 \over \eta}\Big){G_{\rm eff}\over G} 
{\cal D}_2 = 
{2\over 3} (\dot{D_+})^2 -   g{\d G\over G} \Big({\eta-1 \over \eta}\Big)D_+^2,
\label{EOMD2}
\eeq
\end{widetext}
where $D_+$ is the linear growth factor. These equations are straightforward to solve numerically given initial conditions as determined by Eq.~(\ref{F2}) in the large-scale limit, i.e. ${\cal D}_0=17/21$ and ${\cal D}_2=4/21$ with ${\cal D}_i \propto (D_+)^2$. From Eq.~(\ref{delta2}) one can compute the bispectrum using Eq.~(\ref{Btree}), and from it the reduced bispectrum $Q_{123}\equiv B_{123}/(P_1P_2+P_2P_3+P_3P_1)$. Figure~\ref{QDGPGR} shows the result of the ratio of $Q$ to that in standard gravity for $\Omega_m=0.2$ and triangles with $k_1=0.1 \kvecMpc$ and $k_2=2k_1$ as a function of angle $\theta$ between the wavevectors,  showing that there is a few percent enhancement for isosceles configurations, and no correction at all for squeezed triangles ($\theta=0,\pi$) where the kernel in Eq.~(\ref{F2}) vanishes. The $Q$ ratio is roughly independent of scale, and it is most sensitive to  $\Omega_m$, increasing with decreasing $\Omega_m$, again as a result that lower $\Omega_m$ leads to a larger $r_c$ and thus larger $g$ in Eq.~(\ref{GammasLowk}). We compare these results against measurements in N-body simulations in paper~II. 

Although the amplitude of this correction is small compared to the current statistical errors in redshift surveys, the amplitude is model dependent and may be significantly (though only parametrically) different in other modifications of gravity. The characteristic shape dependence (no correction for squeezed triangles, and maximal difference for isosceles triangles) should be robust for the class of models of massive gravity where GR is restored by derivative interactions as in the Vainshtein mechanism. To the extent that such a mechanism is universal~\cite{2006NJPh....8..326D}, we are after a generic feature of modified gravity models that can be tested through the study of cosmological perturbations. In fact, this signature is generic to ``Galileon" models that display a shift symmetry (perhaps remnant of a higher-dimensional Lorentz symmetry) and second-order equations of motion, see~\cite{2008arXiv0811.2197N} where all possible nonlinear operators in 4D with these properties are derived (for recent discussion and extensions of the Galileon model see~\cite{2009PhRvD..79h4003D,2009arXiv0905.1325C,2009arXiv0906.1967D,2009arXiv0905.2943B}). Indeed, it is easy to see that all these operators do posses the same properties as far as higher-order correlators are concerned. That is, using the notation of~\cite{2008arXiv0811.2197N}, a term in the equations of motion 
\beq
{\cal E}_3=( \Box C)^2-(\partial_\mu\partial_\nu C)^2,
\label{E3}
\eeq
gives rise  as we have seen at subhorizon scales to a quadratic kernel
\beq
k_1^2\, k_2^2 - (\k_1\cdot\k_2)^2
\eeq
for the bispectrum that vanishes for squeezed triangles (where $\k_1$ is parallel or antiparallel to $\k_2$). Similarly, the next invariant  
\beq
{\cal E}_4=( \Box C)^3-3( \Box C)(\partial_\mu\partial_\nu C)^2+2\, {\rm Tr}(\partial_\mu\partial_\nu C)^3,
\label{E4}
\eeq
gives rise to a cubic kernel 
\beqa
 k_1^2\, k_2^2\, k_3^2+2(\k_1\cdot\k_2)(\k_2\cdot\k_3)(\k_3\cdot\k_1) & & \nonumber \\
  -k_1^2 (\k_2\cdot\k_3)^2-k_2^2 (\k_3\cdot\k_1)^2-k_3^2 (\k_1\cdot\k_2)^2 & & 
\eeqa
for the trispectrum that vanishes for squeezed trispectrum configurations (where all wavectors are parallel or antiparallel to each other). It's easy to show that the same is true for the last invariant ${\cal E}_5$ as far as the five-point function in Fourier space is concerned.

Finally, it is worth noting that in the normal branch (instead of the self-accelerated branch as we worked so far), the signature in the bispectrum is of opposite sign (suppression for isosceles triangles), since $(\d G/ G)$ changes sign (gravity is stronger than in GR). This should be a generic outcome in models where the extra scalar degree of freedom is not a ghost. For a calculation of the bispectrum and skewness in $f(R)$ models see~\cite{2008arXiv0812.0013B} and~\cite{2008JCAP...09..009T}, respectively.

\subsection{Implications for BAO, Weak Lensing}

We now comment briefly on implications of our results for the Baryon Acoustic Oscillations (BAO) method and weak gravitational lensing as tools to study cosmic acceleration. BAO constraints have been applied to the DGP model~\cite{2006PhRvD..74b3004M,2006ApJ...646....1G,2007PhRvD..75f4003S} with the assumption that nonlinear effects coming from the modified gravity sector would not affect the determination of the acoustic scale done in observations~\cite{2005ApJ...633..560E}. We are now in a better position to assess this assumption.

In GR nonlinearities do shift the acoustic scale compared to linear theory but the effect is small, at the percent level~\cite{2008PhRvD..77b3533C,2008PhRvD..77d3525S,2008ApJ...686...13S,2008arXiv0812.1392K}. The effect can be understood by the contribution of a 90-degree out of phase oscillation by mode-coupling effects~\cite{2008PhRvD..77b3533C}, which develops because nonlinear corrections are a strong function of spectral index~\cite{1996ApJ...473..620S} that is modulated by the linear spectrum BAO's. The bulk of the acoustic scale shift is due to the power spectrum of the second-order density field, Eq.~(\ref{delta2}), in particular the dipole term $\ell=1$ which describes the $({\bf v} \cdot \nabla)$ transport terms in the equation of motion, Eq.~(\ref{EOMdelta}). The $\ell=0,2$ terms contribute to the shift as the mode-coupling power has cross terms of $\ell=1$ with them. Except perhaps for creation of order one higher $\ell$ amplitudes, there are thus two possible changes to the shift in a modified theory of gravity: a change in the $\ell=0,2$ amplitudes, and a change on the $\ell=1$ amplitude itself. 

As we have seen, due to the particular form of the nonlinearities in the DGP model (related to the shift symmetry mentioned above), only $\ell=0,2$ are significantly changed, but from the calculation of the bispectrum, we know that this change is very small, at the few percent level. Therefore, such modification induces changes in the acoustic scale shift by a few percent of a percent, which is negligibly small. The other possible source of concern is a change in the $\ell=1$ amplitude. Such terms are indeed present in this theory, but are highly suppressed. They arise through terms such as $\nabla C \cdot \nabla \Phi$  in the Gauss-Codazzi equation, Eq.~(\ref{Eq44BNLlocal}). It's easy to see that these are of order $(aH/k)^2$ times smaller ($\la 10^{-4} $ at the acoustic scale) than the GR dipole, and that's why we have neglected them.

To summarize, to appreciably change the acoustic scale through nonlinear effects beyond GR, one needs to induce order one changes in the $\ell=0,1,2$ amplitudes of the second-order density field, or equivalently, of the large-scale bispectrum. This is not what happens in this theory, essentially because the Vainshtein scale is so much smaller than the nonlinear scale from standard gravity. In any case, one should be able to correct for nonlinear shifts in any theory of gravity by simultaneously determining  the configuration dependence of the bispectrum with the power spectrum (which can determine the $\ell=0,1,2$ amplitudes), as pointed out in~\cite{2008PhRvD..77b3533C}. 

One caveat needs to be brought up at this point, however. The implicit assumption on this BAO discussion so far is that only mode-coupling power can cause shifts of the linear acoustic scale, but in theories with scale-dependent growth factors (as generic gravity modifications are), another possible source is that the density field propagator (the smoothing kernel that damps linear oscillations in the power spectrum) may contain secondary oscillations, unlike the case in GR where it can be shown to be smooth~\cite{2006PhRvD..73f3520C}. The perceptive reader may have noticed that in Figs.~\ref{GeffNL} and~\ref{DplusR}, $(G_{\rm eff}/G)$ and thus $(D_+^{\rm renorm}/D_+)$ present small amplitude oscillations. These are about $0.2\%$ in amplitude for the power spectrum (a factor of two smaller compared to the 90-degree out of phase oscillations induced by mode-coupling in GR~\cite{2008PhRvD..77b3533C}). At this point it is difficult for us to assess the importance of this, since our calculations are definitely not accurate to that level. Among other things, we relied on the HaloFit fitting formula which has not been calibrated for spectra with BAO's in it. A definite assessment of this  effect will have to await the development of  renormalized perturbation theory for these theories, which is beyond the scope of this paper.

As far as weak lensing at subhorizon scales is concerned, as first pointed out in~\cite{2004PhRvD..69l4015L}, the predictions for the relation of the lensing potential ($\phi_0+\psi_0$, in our convention) to density perturbations $\delta$ in this theory are indistinguishable from GR (see Eq.~\ref{PoissonTrucho}), as the brane-bending mode $C$ does not couple to photons since it is sourced by the trace of the energy momentum tensor and photons are conformal sources (equation of state $w=1/3$) and, in addition, $C$ exactly vanishes on the background solution, see Eq.~(\ref{5Dmetric}).  Therefore, weak lensing in this type of theories is completely unaffected by the extra dynamics in the modified gravity sector, and has no extra information than the density field itself; i.e. the effective Newton's constant driving the lensing potential, Eq.~(\ref{PoissonTrucho}), is not modified. This is not strictly true at the largest scales, where other degrees of freedom enter (through the normal derivatives to the brane), gravity becomes 5D and this is of course followed by the lensing potential~\footnote{In recent work on DGP and degravitation theories, \cite{2009arXiv0903.1292K} obtain lensing modifications at small scales: the reason for this is that they used the linear relation between the lensing and Newtonian potential but the nonlinear relation between the Newtonian potential and the density perturbation.}, see e.g.~\cite{2001PhRvD..64h3004U,2004astro.ph..9224B,2009arXiv0903.1292K}, this may be again described by a running of the gravitational constant appearing in Eq.~(\ref{PoissonTrucho}), which ignored normal derivatives.

\section{Conclusions}

In this paper, we have studied linear and nonlinear solutions for cosmological perturbations in brane-induced gravity in five dimensions. Our results provide improved understanding of the linear solutions and derive new predictions at the nonlinear level that should be also relevant for other theories of modified gravity with similar characteristics.  Most of the predictions developed here are compared to numerical simulations in paper~II, finding good agreement. Given the length of this paper, it seems appropriate here to summarize  the main results, discuss their validity and suggest possible improvements.


We solved for the linear evolution of perturbations taking advantage of the symmetry under gauge transformations along the extra-dimension to decouple the bulk equations in the quasistatic approximation. This is a different route than the popular approach in the literature of using the master equation, and provides in our view a more physically transparent procedure (to the extent that five-dimensional physics can be so!).  

We have checked the validity of the quasistatic approximation for all metric perturbations, and pointed out that the derivatives normal to the brane belong to the quasistatic approximation, giving rise to 5D behavior in the static on-brane potentials, in close analogy to the graviton propagator. In the past, these contributions have been considered as deviations from the quasistatic approximation, e.g.~\cite{2007PhRvD..75f4002S,2008PhRvD..77h3512C}; therefore, the validity of this approximation for cosmological perturbations is likely better than thought before. In addition, it is worth emphasizing that these normal derivatives induce, from the 4D point of view, unusual nonlocal operators (see Eq.~\ref{sqrtnabla}), this cannot be obtained by a local 4D theory no matter how complicated, thus making brane-induced gravity distinct from arbitrary (but local) dark energy models. A similar regime is expected in theories with more extra dimensions, as gravity ``cascades down" from higher to lower dimensionality as smaller distances are probed.


Linear theory can only be, however, a limited description of the physics in the modified gravity sector, because if valid at all scales the theory would be ruled out by local experiments, e.g. by showing that light deflection and nonrelativistic motion display different gravitational constants. Nonlinear effects, through the Vainshtein mechanism, provide a natural way out of this situation. We studied the nonlinearities in the bulk and brane equations, and justified the approximations made, which form the basis for carrying out numerical simulations of structure formation (see paper~II). Along the way, we justified why the ghost-like brane-bending mode interacting with healthy modes does not cause catastrophic instabilities for cosmological solutions. 

We studied in detail the Vainshtein mechanism, by which the theory becomes GR at small scales. In this regard, it is useful to think geometrically, in terms of the extrinsic curvature $K$ of the brane; after all, cosmic acceleration is driven in this theory by a balance between intrinsic (Ricci, $R$) and extrinsic curvature. At the level of perturbations, and within the quasistatic approximation, there are three distinct regimes:

\begin{enumerate}

\item {\em 5D regime}, with $\delta K/\bar{K} \sim \delta R/\bar{R}$ and $\delta K$ dominated by normal derivatives, giving rise to 5D behavior and a scale-dependent renormalization of $G$,

\item {\em scalar-tensor regime}, with $\delta K/\bar{K} \sim \delta R/\bar{R}$ and $\delta K$ dominated by the brane bending mode $C$, leading to scale-independent (but time dependent) renormalization of the gravitational constant,

\item {\em Vainshtein regime}, with $\delta K/\bar{K} \ll \delta R/\bar{R}$ and $\delta K$ dominated by the brane bending mode $C$, but subdominant compared to density fluctuations ($\delta K/\bar{K} \sim \delta^{1/2}$), thus its effect on the physics at the brane is greatly reduced, $G_{\rm eff}/G$ becomes scale-dependent through nonlinear effects, approaching unity at small scales where GR is recovered.

\end{enumerate}
These three stages can be seen in Fig.~\ref{GeffNL}, although the second stage is greatly reduced at low redshift.

We showed that throughout evolution, the weak field approximation is always obeyed (this is also supported by numerical simulations in paper~II), the quantity that goes nonlinear in the Vainshtein mechanism is the extrinsic curvature fluctuation $\delta K/\bar{K}$. We also showed that the condition $\delta K/\bar{K}=1$ can be written in the familiar way ($r_*^3=r_g\,r_c^2$) for the Vainshtein scale of isolated perturbations, making the connection with the standard Schwarzschild solution. 

We calculated the impact of the Vainshtein mechanism on the nonlinear density power spectrum. To do so, we derived an explicit modified Poisson equation that relates the Newtonian potential $\phi_0$ to density perturbations $\delta$, Eq.~(\ref{NLPoisson}). We showed that this nonlinear integral equation can be rewritten as an effective linear plus quadratic response for $\phi_0$ as a function of $\delta$ (Eq.~\ref{PoissonGamma}) by introducing the two- and three-point  propagators of the Poisson equation. The two-point propagator describes the running of the gravitational constant due to the Vainshtein mechanism, and  in order to recover the right asymptotics at small scales (see Eq.~\ref{GammaAsymp}) one has to resum it. The three-point propagator only contributes at large scales, and gives rise to new signatures of modified gravity as corrections to non-Gaussianity. 

To resum the two-point modified Poisson propagator one has to solve a complicated integral equation, which involves the density correlators we are interested in; we implemented an iterative procedure based on knowledge of the power spectrum and bispectrum in GR with the same expansion history. Our method can be improved by recomputing the power spectrum and bispectrum at each step given the previous iteration solution of the two-point propagator, for self-consistency. Another aspect that may merit improvement is the development of better GR fitting functions for the power spectrum and bispectrum. To put our resummation in a more rigorous footing, one needs to develop in full e.g. renormalized perturbation theory~\cite{2006PhRvD..73f3519C} for modified gravity theories (also needed for checking on additional BAO systematics), which is left for future work. 

After resummation of the two-point propagator is achieved, the effective equations of motion for density and velocity fields is remarkably similar to standard gravity, see Eq.~(\ref{EOMdelta}). The difference is a slowly varying gravitational constant, and a new vertex that gives rise to non-Gaussianity. Neglecting the latter (justified by the small corrections induced in the bispectrum) we showed that to calculate the nonlinear power spectrum, one can implement the GR fitting formula for the power spectrum on a renormalized linear spectrum that already contains the Vainshtein mechanism through the renormalization of $G$. This renormalized linear spectrum can also be used for predicting the mass function of dark matter halos (see paper~II). The calculated nonlinear power spectrum (Fig.~\ref{PowerDGP}) approaches that of GR with the same expansion history, as expected, but the transition is rather gentle. See paper~II for comparison of these predictions with N-body simulations. 
 
 It is worth noting that at the level of the power spectrum, and to a reasonably good approximation, the effect of nonlinearities in the modified gravity sector may be absorbed into a renormalization of the gravitational constant. Since the relation between the lensing potential and density perturbations is entirely unaffected by the extra physics in these theories, the modified gravity can be described in this approximation by a single function, an effective gravitational constant $G_{\rm eff}(k,a)$ for non-relativistic motion that depends on space and time; the caveat is that it cannot be computed from linear theory. This simple result (checked against simulations in paper~II)  may have important practical consequences for parametrizing modified gravity in observations.
 
The extra nonlinear vertex in Eq.~(\ref{EOMdelta}) gives rise to additional non-Gaussianity, a signature of the Vainshtein mechanism at large-scales. We calculated the corrections induced by it in the bispectrum, and while the amplitude is small and model dependent, the characteristic shape dependence (vanishing for squeezed triangles) is generic to modified gravity theories that are of ``Galileon" type~\cite{2008arXiv0811.2197N}, i.e. where the dynamics of the additional scalar degree of freedom is invariant under shifts of its gradient. We pointed out that similar results hold for higher-order correlators than the bispectrum in these type of theories. These signatures of nonlinearities in modified gravity theories can be used as a diagnostic in future observations.

\acknowledgements 

We thank R.~Crittenden, G.~Dvali, S.~Dubovsky, J.~Frieman, J.~Garriga, A.~Gruzinov, W.~Hu, L.~Hui, J.~ Khoury, A.~Iglesias, B.~Jain, M.~Kleban, A.~Lue, A.~Nicolis,  M.~Porrati, O.~Pujolas,  I.~Sawicki, F.~Schmidt, R.~Sheth, G.~Starkman, and especially C.~Deffayet and G.~Gabadadze for many useful discussions. We thank K.C. Chan for his patience in holding paper~II for several months while this paper was being typed. 

We thank the Aspen Center for Physics, the Perimeter Institute, the Center for Cosmological Physics at the University of Chicago, and the Department of Astronomy and Physics at the University of Pennsylvania for hospitality while this work was being done. We also thank the participants of the workshops on Infrared Modification of Gravity (Perimeter Institute), Consistent Infrared Modification of Gravity (Paris) and Testing General Relativity in the Cosmos (Aspen) for feedback on this work. 

This research was partially supported by grants NSF AST-0607747 and NASA NNG06GH21G.

%

\bibliography{masterbiblio}

\begin{thebibliography}{100}

\bibitem{2004PhRvD..69d4005L}
A.~{Lue}, R.~{Scoccimarro}, and G.~{Starkman}.
\newblock {Differentiating between modified gravity and dark energy}.
\newblock {\em \prd}, 69(4):044005--+, February 2004.

\bibitem{2001PhRvD..64h3004U}
J.-P. {Uzan} and F.~{Bernardeau}.
\newblock {Lensing at cosmological scales: A test of higher dimensional
  gravity}.
\newblock {\em \prd}, 64(8):083004--+, October 2001.

\bibitem{2004astro.ph..9224B}
F.~{Bernardeau}.
\newblock {Constraints on higher-dimensional gravity from the cosmic shear
  three-point correlation function}.
\newblock {\em ArXiv Astrophysics e-prints}, September 2004.

\bibitem{2005PhRvD..71h3004S}
C.~{Sealfon}, L.~{Verde}, and R.~{Jimenez}.
\newblock {Limits on deviations from the inverse-square law on megaparsec
  scales}.
\newblock {\em \prd}, 71(8):083004--+, April 2005.

\bibitem{2005PhRvD..71b4026S}
Y.-S. {Song}.
\newblock {Looking for an extra dimension with tomographic cosmic shear}.
\newblock {\em \prd}, 71(2):024026--+, January 2005.

\bibitem{2006ApJ...648..797B}
E.~{Bertschinger}.
\newblock {On the Growth of Perturbations as a Test of Dark Energy and
  Gravity}.
\newblock {\em \apj}, 648:797--806, September 2006.

\bibitem{2006PhRvD..74b3512K}
L.~{Knox}, Y.-S. {Song}, and J.~A. {Tyson}.
\newblock {Distance-redshift and growth-redshift relations as two windows on
  acceleration and gravitation: Dark energy or new gravity?}
\newblock {\em \prd}, 74(2):023512--+, July 2006.

\bibitem{2006PhRvD..74d3513I}
M.~{Ishak}, A.~{Upadhye}, and D.~N. {Spergel}.
\newblock {Probing cosmic acceleration beyond the equation of state:
  Distinguishing between dark energy and modified gravity models}.
\newblock {\em \prd}, 74(4):043513--+, August 2006.

\bibitem{2007PhRvD..75b3519H}
D.~{Huterer} and E.~V. {Linder}.
\newblock {Separating dark physics from physical darkness: Minimalist modified
  gravity versus dark energy}.
\newblock {\em \prd}, 75(2):023519--+, January 2007.

\bibitem{2007PhRvL..98l1301K}
M.~{Kunz} and D.~{Sapone}.
\newblock {Dark Energy versus Modified Gravity}.
\newblock {\em Physical Review Letters}, 98(12):121301--+, March 2007.

\bibitem{2007PhRvD..76b3507C}
R.~{Caldwell}, A.~{Cooray}, and A.~{Melchiorri}.
\newblock {Constraints on a new post-general relativity cosmological
  parameter}.
\newblock {\em \prd}, 76(2):023507--+, July 2007.

\bibitem{2007astro.ph..1782F}
P.~{Fosalba} and O.~{Dor{\'e}}.
\newblock {Probing the Largest Cosmological Scales with the CMB-Velocity
  Correlation}.
\newblock {\em ArXiv Astrophysics e-prints}, January 2007.

\bibitem{2008PhRvD..77j3513D}
S.~F. {Daniel}, R.~R. {Caldwell}, A.~{Cooray}, and A.~{Melchiorri}.
\newblock {Large scale structure as a probe of gravitational slip}.
\newblock {\em \prd}, 77(10):103513--+, May 2008.

\bibitem{2008PhRvD..78b4015B}
E.~{Bertschinger} and P.~{Zukin}.
\newblock {Distinguishing modified gravity from dark energy}.
\newblock {\em \prd}, 78(2):024015--+, July 2008.

\bibitem{2008PhRvD..78j3509F}
W.~{Fang}, S.~{Wang}, W.~{Hu}, Z.~{Haiman}, L.~{Hui}, and M.~{May}.
\newblock {Challenges to the DGP model from horizon-scale growth and geometry}.
\newblock {\em \prd}, 78(10):103509--+, November 2008.

\bibitem{2008PhRvD..78f3503J}
B.~{Jain} and P.~{Zhang}.
\newblock {Observational tests of modified gravity}.
\newblock {\em \prd}, 78(6):063503--+, September 2008.

\bibitem{2008PhRvD..78d3514A}
V.~{Acquaviva}, A.~{Hajian}, D.~N. {Spergel}, and S.~{Das}.
\newblock {Next generation redshift surveys and the origin of cosmic
  acceleration}.
\newblock {\em \prd}, 78(4):043514--+, August 2008.

\bibitem{2008PhRvD..78d3002S}
F.~{Schmidt}.
\newblock {Weak lensing probes of modified gravity}.
\newblock {\em \prd}, 78(4):043002--+, August 2008.

\bibitem{2008MNRAS.390..131A}
M.~A. {Amin}, R.~V. {Wagoner}, and R.~D. {Blandford}.
\newblock {A subhorizon framework for probing the relationship between the
  cosmological matter distribution and metric perturbations}.
\newblock {\em \mnras}, 390:131--142, October 2008.

\bibitem{2008arXiv0807.0810S}
Y.-S. {Song} and W.~J. {Percival}.
\newblock {Reconstructing the History of Structure Formation using Peculiar
  Velocities}.
\newblock {\em ArXiv e-prints}, July 2008.

\bibitem{2008arXiv0809.3791Z}
G.-B. {Zhao}, L.~{Pogosian}, A.~{Silvestri}, and J.~{Zylberberg}.
\newblock {Searching for modified growth patterns with tomographic surveys}.
\newblock {\em ArXiv e-prints}, September 2008.

\bibitem{2008arXiv0812.0002S}
Y.-S. {Song} and O.~{Dor{\'e}}.
\newblock {Testing gravity using weak gravitational lensing and redshift
  surveys}.
\newblock {\em ArXiv e-prints}, December 2008.

\bibitem{2008arXiv0812.2244A}
N.~{Afshordi}, G.~{Geshnizjani}, and J.~{Khoury}.
\newblock {Observational Evidence for Cosmological-Scale Extra Dimensions}.
\newblock {\em ArXiv e-prints}, December 2008.

\bibitem{2009ApJ...690..923Z}
H.~{Zhan}, L.~{Knox}, and J.~A. {Tyson}.
\newblock {Distance, Growth Factor, and Dark Energy Constraints from
  Photometric Baryon Acoustic Oscillation and Weak Lensing Measurements}.
\newblock {\em \apj}, 690:923--936, January 2009.

\bibitem{2009JCAP...01..048S}
Y.-S. {Song} and K.~{Koyama}.
\newblock {Consistency test of general relativity from large scale structure of
  the universe}.
\newblock {\em Journal of Cosmology and Astro-Particle Physics}, 1:48--+,
  January 2009.

\bibitem{2009arXiv0901.0919D}
S.~F. {Daniel}, R.~R. {Caldwell}, A.~{Cooray}, P.~{Serra}, and A.~{Melchiorri}.
\newblock {A Multi-Parameter Investigation of Gravitational Slip}.
\newblock {\em ArXiv e-prints}, January 2009.

\bibitem{1955PhRv...98.1118K}
R.~H. {Kraichnan}.
\newblock {Special-Relativistic Derivation of Generally Covariant Gravitation
  Theory}.
\newblock {\em Physical Review}, 98:1118--1122, May 1955.

\bibitem{1970GReGr...1....9D}
S.~{Deser}.
\newblock {Self-interaction and gauge invariance}.
\newblock {\em General Relativity and Gravitation}, 1:9--18, March 1970.

\bibitem{1975AnPhy..89..193B}
D.~G. {Boulware} and S.~{Deser}.
\newblock {Classical general relativity derived from quantum gravity}.
\newblock {\em Annals of Physics}, 89:193--240, January 1975.

\bibitem{1986PhRvD..33.3613W}
R.~M. {Wald}.
\newblock {Spin-two fields and general covariance}.
\newblock {\em \prd}, 33:3613--3625, June 1986.

\bibitem{2003flog.book.....F}
R.~P. {Feynman}, F.~B. {Mor{\'{\i}}nigo}, W.~G. {Wagner}, and B.~{Hatfield}.
\newblock {\em {Feynman lectures on gravitation}}.
\newblock Westview Press, 2003.

\bibitem{2004PhRvD..70d3528C}
S.~M. {Carroll}, V.~{Duvvuri}, M.~{Trodden}, and M.~S. {Turner}.
\newblock {Is cosmic speed-up due to new gravitational physics?}
\newblock {\em \prd}, 70(4):043528--+, August 2004.

\bibitem{2007PhRvD..76f4004H}
W.~{Hu} and I.~{Sawicki}.
\newblock {Models of f(R) cosmic acceleration that evade solar system tests}.
\newblock {\em \prd}, 76(6):064004--+, September 2007.

\bibitem{2008arXiv0811.2197N}
A.~{Nicolis}, R.~{Rattazzi}, and E.~{Trincherini}.
\newblock {The galileon as a local modification of gravity}.
\newblock {\em ArXiv e-prints}, November 2008.

\bibitem{2000PhLB..485..208D}
G.~{Dvali}, G.~{Gabadadze}, and M.~{Porrati}.
\newblock {4D gravity on a brane in 5D Minkowski space}.
\newblock {\em Physics Letters B}, 485:208--214, 2000.

\bibitem{2007JHEP...05..045K}
N.~{Kaloper} and D.~{Kiley}.
\newblock {Charting the landscape of modified gravity}.
\newblock {\em Journal of High Energy Physics}, 5:45--045, May 2007.

\bibitem{2008PhRvL.100y1603D}
C.~{de Rham}, G.~{Dvali}, S.~{Hofmann}, J.~{Khoury}, O.~{Pujol{\`a}s},
  M.~{Redi}, and A.~J. {Tolley}.
\newblock {Cascading Gravity: Extending the Dvali-Gabadadze-Porrati Model to
  Higher Dimension}.
\newblock {\em Physical Review Letters}, 100(25):251603--+, June 2008.

\bibitem{2008JCAP...02..011D}
C.~{de Rham}, S.~{Hofmann}, J.~{Khoury}, and A.~J. {Tolley}.
\newblock {Cascading gravity and degravitation}.
\newblock {\em Journal of Cosmology and Astro-Particle Physics}, 2:11--+,
  February 2008.

\bibitem{2007PhRvD..76h4006D}
G.~{Dvali}, S.~{Hofmann}, and J.~{Khoury}.
\newblock {Degravitation of the cosmological constant and graviton width}.
\newblock {\em \prd}, 76(8):084006--+, October 2007.

\bibitem{2002PhRvD..66j4025D}
T.~{Damour}, I.{}I. {Kogan}, and A.~{Papazoglou}.
\newblock {Nonlinear bigravity and cosmic acceleration}.
\newblock {\em \prd}, 66:104025, 2002.

\bibitem{2007PhRvD..76j4036B}
D.~{Blas}, C.~{Deffayet}, and J.~{Garriga}.
\newblock {Bigravity and Lorentz-violating massive gravity}.
\newblock {\em \prd}, 76(10):104036--+, November 2007.

\bibitem{2007NuPhS.171...88G}
G.~{Gabadadze}.
\newblock {Carg{\`e}se Lectures on Brane Induced Gravity}.
\newblock {\em Nuclear Physics B Proceedings Supplements}, 171:88--98,
  September 2007.

\bibitem{1972PhLB...39..393V}
A.~I. {Vainshtein}.
\newblock {To the problem of nonvanishing gravitation mass}.
\newblock {\em Physics Letters B}, 39:393--394, May 1972.

\bibitem{2002PhRvD..65d4026D}
C.~{Deffayet}, G.~{Dvali}, G.~{Gabadadze}, and A.~{Vainshtein}.
\newblock {Nonperturbative continuity in graviton mass versus perturbative
  discontinuity}.
\newblock {\em \prd}, 65(4):044026--+, 2002.

\bibitem{2006NJPh....8..326D}
G.~{Dvali}.
\newblock {Predictive power of strong coupling in theories with large distance
  modified gravity}.
\newblock {\em New Journal of Physics}, 8:326--+, December 2006.

\bibitem{2009arXiv0901.0393B}
E.~{Babichev}, C.~{Deffayet}, and R.~{Ziour}.
\newblock {The Vainshtein mechanism in the Decoupling Limit of massive
  gravity}.
\newblock {\em ArXiv e-prints}, January 2009.

\bibitem{2004PhRvD..69l4015L}
A.~{Lue}, R.~{Scoccimarro}, and G.~D. {Starkman}.
\newblock {Probing Newton's constant on vast scales: Dvali-Gabadadze-Porrati
  gravity, cosmic acceleration, and large scale structure}.
\newblock {\em \prd}, 69(12):124015--+, June 2004.

\bibitem{LSSbrane2}
K.~C. {Chan} and R.~{Scoccimarro}.
\newblock Large-scale structure in brane-induced gravity. ii. numerical
  simulations.
\newblock {\em arXiv:0906.4548}, 2009.

\bibitem{2004JHEP...05..074H}
N.~A. {Hamed}, H.~S. {Cheng}, M.~A. {Luty}, and S.~{Mukohyama}.
\newblock {Ghost Condensation and a Consistent IR Modification of Gravity}.
\newblock {\em Journal of High Energy Physics}, 5:74--+, May 2004.

\bibitem{2004JHEP...10..076D}
S.~L. {Dubovsky}.
\newblock {Phases of massive gravity}.
\newblock {\em Journal of High Energy Physics}, 10:76--+, October 2004.

\bibitem{2001PhLB..502..199D}
C.~{Deffayet}.
\newblock {Cosmology on a brane in Minkowski bulk}.
\newblock {\em Physics Letters B}, 502:199--208, March 2001.

\bibitem{2002PhRvD..65d4023D}
C.~{Deffayet}, G.~{Dvali}, and G.~{Gabadadze}.
\newblock {Accelerated universe from gravity leaking to extra dimensions}.
\newblock {\em \prd}, 65(4):044023--+, February 2002.

\bibitem{2003JHEP...09..029L}
M.~A. {Luty}, M.~{Porrati}, and R.~{Rattazzi}.
\newblock {Strong interactions and stability in the DGP model}.
\newblock {\em Journal of High Energy Physics}, 9:29--+, September 2003.

\bibitem{2004JHEP...06..059N}
A.~{Nicolis} and R.~{Rattazzi}.
\newblock {Classical and Quantum Consistency of the DGP Model}.
\newblock {\em Journal of High Energy Physics}, 6:59--+, June 2004.

\bibitem{2006JCAP...08..012D}
C.~{Deffayet}, G.~{Gabadadze}, and A.~{Iglesias}.
\newblock {Perturbations of the self-accelerated Universe}.
\newblock {\em Journal of Cosmology and Astro-Particle Physics}, 8:12--+,
  August 2006.

\bibitem{2006PhRvD..73d4016G}
D.~{Gorbunov}, K.~{Koyama}, and S.~{Sibiryakov}.
\newblock {More on ghosts in the Dvali-Gabadaze-Porrati model}.
\newblock {\em \prd}, 73(4):044016--+, February 2006.

\bibitem{2006JHEP...10..066C}
C.~{Charmousis}, R.~{Gregory}, N.~{Kaloper}, and A.~{Padilla}.
\newblock {DGP specteroscopy}.
\newblock {\em Journal of High Energy Physics}, 10:66--+, October 2006.

\bibitem{2007PhRvD..76j4041I}
K.~{Izumi}, K.~{Koyama}, O.~{Pujol{\`a}s}, and T.~{Tanaka}.
\newblock {Bubbles in the self-accelerating universe}.
\newblock {\em \prd}, 76(10):104041--+, November 2007.

\bibitem{2002PhRvD..66b4019D}
C.~{Deffayet}, S.~?J. {Landau}, J.~{Raux}, M.~{Zaldarriaga}, and P.~{Astier}.
\newblock {Supernovae, CMB, and gravitational leakage into extra dimensions}.
\newblock {\em \prd}, 66(2):024019--+, 2002.

\bibitem{2006PhRvD..74b3004M}
R.~{Maartens} and E.~{Majerotto}.
\newblock {Observational constraints on self-accelerating cosmology}.
\newblock {\em \prd}, 74(2):023004--+, July 2006.

\bibitem{2007PhRvD..75f4003S}
Y.-S. {Song}, I.~{Sawicki}, and W.~{Hu}.
\newblock {Large-scale tests of the Dvali-Gabadadze-Porrati model}.
\newblock {\em \prd}, 75(6):064003--+, March 2007.

\bibitem{2007PhRvD..76f3503W}
S.~{Wang}, L.~{Hui}, M.~{May}, and Z.~{Haiman}.
\newblock {Is modified gravity required by observations? An empirical
  consistency test of dark energy models}.
\newblock {\em \prd}, 76(6):063503--+, September 2007.

\bibitem{2006JCAP...01..016K}
K.~{Koyama} and R.~{Maartens}.
\newblock {Structure formation in the Dvali Gabadadze Porrati cosmological
  model}.
\newblock {\em Journal of Cosmology and Astro-Particle Physics}, 1:16--+,
  January 2006.

\bibitem{2007PhRvD..75f4002S}
I.~{Sawicki}, Y.-S. {Song}, and W.~{Hu}.
\newblock {Near-horizon solution for Dvali-Gabadadze-Porrati perturbations}.
\newblock {\em \prd}, 75(6):064002--+, March 2007.

\bibitem{2008PhRvD..77h3512C}
A.~{Cardoso}, K.~{Koyama}, S.~S. {Seahra}, and F.~P. {Silva}.
\newblock {Cosmological perturbations in the DGP braneworld: Numeric solution}.
\newblock {\em \prd}, 77(8):083512--+, April 2008.

\bibitem{2000PhRvD..62h4015M}
S.~{Mukohyama}.
\newblock {Gauge-invariant gravitational perturbations of maximally symmetric
  spacetimes}.
\newblock {\em \prd}, 62(8):084015--+, October 2000.

\bibitem{2002PhRvD..66j3504D}
C.~{Deffayet}.
\newblock {On brane world cosmological perturbations}.
\newblock {\em \prd}, 66(10):103504--+, November 2002.

\bibitem{2009arXiv0905.0858S}
F.~{Schmidt}.
\newblock {Self-Consistent Cosmological Simulations of DGP Braneworld Gravity}.
\newblock {\em ArXiv e-prints}, May 2009.

\bibitem{2000PhRvD..62f4022K}
H.~{Kodama}, A.~{Ishibashi}, and O.~{Seto}.
\newblock {Brane world cosmology: Gauge-invariant formalism for perturbation}.
\newblock {\em \prd}, 62(6):064022--+, September 2000.

\bibitem{2002hep.th....5220R}
A.~{Riazuelo}, F.~{Vernizzi}, D.~{Steer}, and R.~{Durrer}.
\newblock {Gauge invariant cosmological perturbation theory for braneworlds}.
\newblock {\em ArXiv High Energy Physics - Theory e-prints}, May 2002.

\bibitem{1984PThPS..78....1K}
H.~{Kodama} and M.~{Sasaki}.
\newblock {Cosmological Perturbation Theory}.
\newblock {\em Progress of Theoretical Physics Supplement}, 78:1--+, 1984.

\bibitem{1992PhR...215..203M}
V.~F. {Mukhanov}, H.~A. {Feldman}, and R.~H. {Brandenberger}.
\newblock {Theory of cosmological perturbations}.
\newblock {\em \physrep}, 215:203--333, June 1992.

\bibitem{2000PhRvL..84.2778G}
J.~{Garriga} and T.~{Tanaka}.
\newblock {Gravity in the Randall-Sundrum Brane World}.
\newblock {\em Physical Review Letters}, 84:2778--2781, March 2000.

\bibitem{2008GReGr..40.1997A}
R.~{Arnowitt}, S.~{Deser}, and C.~W. {Misner}.
\newblock {Republication of: The dynamics of general relativity}.
\newblock {\em General Relativity and Gravitation}, 40:1997--2027, September
  2008.

\bibitem{2000PhLB..484..112D}
G.~{Dvali}, G.~{Gabadadze}, and M.~{Porrati}.
\newblock {Metastable gravitons and infinite volume extra dimensions}.
\newblock {\em Physics Letters B}, 484:112--118, 2000.

\bibitem{2002PhRvD..65b4031D}
G.~{Dvali}, G.~{Gabadadze}, M.~{Kolanovi{\'c}}, and F.~{Nitti}.
\newblock {Scales of gravity}.
\newblock {\em \prd}, 65(2):024031--+, 2002.

\bibitem{1970PhRvD...2.2255I}
Y.~{Iwasaki}.
\newblock {Consistency Condition for Propagators}.
\newblock {\em \prd}, 2:2255--2256, November 1970.

\bibitem{1970NuPhB..22..397V}
H.~{van Dam} and M.~{Veltman}.
\newblock {Massive and mass-less Yang-Mills and gravitational fields}.
\newblock {\em Nuclear Physics B}, 22:397--411, September 1970.

\bibitem{1970JETPL..12..312Z}
V.~I. {Zakharov}.
\newblock {Linearized Gravitation Theory and the Graviton Mass}.
\newblock {\em Soviet Journal of Experimental and Theoretical Physics Letters},
  12:312--+, 1970.

\bibitem{2007PhRvD..75h4040K}
K.~{Koyama} and F.~P. {Silva}.
\newblock {Nonlinear interactions in a cosmological background in the
  Dvali-Gabadadze-Porrati braneworld}.
\newblock {\em \prd}, 75(8):084040--+, April 2007.

\bibitem{2004PhRvD..69b4001T}
T.~{Tanaka}.
\newblock {Weak gravity in the Dvali-Gabadadze-Porrati braneworld model}.
\newblock {\em \prd}, 69(2):024001--+, January 2004.

\bibitem{2003AnPhy.305...96A}
N.~{Arkani-Hamed}, H.~{Georgi}, and M.~D. {Schwartz}.
\newblock {Effective field theory for massive gravitons and gravity in theory
  space}.
\newblock {\em Annals of Physics}, 305:96--118, June 2003.

\bibitem{2003PhRvD..68b3509C}
S.~M. {Carroll}, M.~{Hoffman}, and M.~{Trodden}.
\newblock {Can the dark energy equation-of-state parameter w be less than -1?}
\newblock {\em \prd}, 68(2):023509--+, July 2003.

\bibitem{2004PhRvD..70d3543C}
J.~M. {Cline}, S.~{Jeon}, and G.~D. {Moore}.
\newblock {The phantom menaced: Constraints on low-energy effective ghosts}.
\newblock {\em \prd}, 70(4):043543--+, August 2004.

\bibitem{2002PhRvD..66d3509L}
A.~{Lue}.
\newblock {Cosmic strings in a braneworld theory with metastable gravitons}.
\newblock {\em \prd}, 66(4):043509--+, August 2002.

\bibitem{2002PhLB..534..209P}
M.~{Porrati}.
\newblock {Fully covariant van Dam-Veltman-Zakharov discontinuity, and absence
  thereof}.
\newblock {\em Physics Letters B}, 534:209--215, 2002.

\bibitem{2005NewA...10..311G}
A.~{Gruzinov}.
\newblock {On the graviton mass}.
\newblock {\em New Astronomy}, 10:311--314, March 2005.

\bibitem{2005PhRvD..72h4024G}
G.~{Gabadadze} and A.~{Iglesias}.
\newblock {Schwarzschild solution in brane induced gravity}.
\newblock {\em \prd}, 72(8):084024--+, October 2005.

\bibitem{2008CQGra..25o4008G}
G.~{Gabadadze} and A.~{Iglesias}.
\newblock {Mass screening in modified gravity}.
\newblock {\em Classical and Quantum Gravity}, 25(15):154008--+, August 2008.

\bibitem{2009arXiv0905.2966H}
L.~{Hui}, A.~{Nicolis}, and C.~{Stubbs}.
\newblock {Equivalence Principle Implications of Modified Gravity Models}.
\newblock {\em ArXiv e-prints}, May 2009.

\bibitem{2009arXiv0903.1292K}
J.~{Khoury} and M.~{Wyman}.
\newblock {N-Body Simulations of DGP and Degravitation Theories}.
\newblock {\em ArXiv e-prints}, March 2009.

\bibitem{2006PhRvD..73f3519C}
M.~{Crocce} and R.~{Scoccimarro}.
\newblock {Renormalized cosmological perturbation theory}.
\newblock {\em \prd}, 73(6):063519--+, March 2006.

\bibitem{2008PhRvD..78j3521B}
F.~{Bernardeau}, M.~{Crocce}, and R.~{Scoccimarro}.
\newblock {Multipoint propagators in cosmological gravitational instability}.
\newblock {\em \prd}, 78(10):103521--+, November 2008.

\bibitem{2003MNRAS.341.1311S}
R.{}E. {Smith}, J.{}A. {Peacock}, A.~{Jenkins}, S.{}D.{}M. {White}, C.{}S.
  {Frenk}, F.{}R. {Pearce}, P.{}A. {Thomas}, G.~{Efstathiou}, and H.{}M.{}P.
  {Couchman}.
\newblock {Stable clustering, the halo model and non-linear cosmological power
  spectra}.
\newblock {\em \mnras}, 341:1311--1332, 2003.

\bibitem{2001MNRAS.325.1312S}
R.~{Scoccimarro} and H.{}M.{}P. {Couchman}.
\newblock {A fitting formula for the non-linear evolution of the bispectrum}.
\newblock {\em \mnras}, 325:1312--1316, 2001.

\bibitem{1999ApJ...520...35S}
R.~{Scoccimarro} and J.{}A. {Frieman}.
\newblock {Hyperextended Cosmological Perturbation Theory: Predicting Nonlinear
  Clustering Amplitudes}.
\newblock {\em \apj}, 520:35--44, 1999.

\bibitem{2009arXiv0902.0618K}
K.~{Koyama}, A.~{Taruya}, and T.~{Hiramatsu}.
\newblock {Non-linear Evolution of Matter Power Spectrum in Modified Theory of
  Gravity}.
\newblock {\em ArXiv e-prints}, February 2009.

\bibitem{2005PhRvD..72d3529L}
E.~V. {Linder}.
\newblock {Cosmic growth history and expansion history}.
\newblock {\em \prd}, 72(4):043529--+, August 2005.

\bibitem{2008PhRvD..78l3524O}
H.~{Oyaizu}, M.~{Lima}, and W.~{Hu}.
\newblock {Nonlinear evolution of f(R) cosmologies. II. Power spectrum}.
\newblock {\em \prd}, 78(12):123524--+, December 2008.

\bibitem{2007PhRvD..76j4043H}
W.~{Hu} and I.~{Sawicki}.
\newblock {Parametrized post-Friedmann framework for modified gravity}.
\newblock {\em \prd}, 76(10):104043--+, November 2007.

\bibitem{2009arXiv0905.1325C}
N.~{Chow} and J.~{Khoury}.
\newblock {Galileon Cosmology}.
\newblock {\em ArXiv e-prints}, May 2009.

\bibitem{2009PhRvD..79h4003D}
C.~{Deffayet}, G.~{Esposito-Far{\`e}se}, and A.~{Vikman}.
\newblock {Covariant Galileon}.
\newblock {\em \prd}, 79(8):084003--+, April 2009.

\bibitem{2009arXiv0906.1967D}
C.~{Deffayet}, S.~{Deser}, and G.~{Esposito-Farese}.
\newblock {Generalized Galileons: All scalar models whose curved background
  extensions maintain second-order field equations and stress-tensors}.
\newblock {\em ArXiv e-prints}, June 2009.

\bibitem{2009arXiv0905.2943B}
E.~{Babichev}, C.~{Deffayet}, and R.~{Ziour}.
\newblock {k-Mouflage gravity}.
\newblock {\em ArXiv e-prints}, May 2009.

\bibitem{2008arXiv0812.0013B}
A.~{Borisov} and B.~{Jain}.
\newblock {Three-Point Correlations in f(R) Models of Gravity}.
\newblock {\em ArXiv e-prints}, December 2008.

\bibitem{2008JCAP...09..009T}
T.~{Tatekawa} and S.~{Tsujikawa}.
\newblock {Second-order matter density perturbations and skewness in scalar
  tensor modified gravity models}.
\newblock {\em Journal of Cosmology and Astro-Particle Physics}, 9:9--+,
  September 2008.

\bibitem{2006ApJ...646....1G}
Z.-K. {Guo}, Z.-H. {Zhu}, J.~S. {Alcaniz}, and Y.-Z. {Zhang}.
\newblock {Constraints on the Dvali-Gabadadze-Porrati Model from Recent
  Supernova Observations and Baryon Acoustic Oscillations}.
\newblock {\em \apj}, 646:1--7, July 2006.

\bibitem{2005ApJ...633..560E}
D.~J. {Eisenstein}, I.~{Zehavi}, D.~W. {Hogg}, R.~{Scoccimarro}, M.~R.
  {Blanton}, R.~C. {Nichol}, R.~{Scranton}, H.-J. {Seo}, M.~{Tegmark},
  Z.~{Zheng}, S.~F. {Anderson}, J.~{Annis}, N.~{Bahcall}, J.~{Brinkmann},
  S.~{Burles}, F.~J. {Castander}, A.~{Connolly}, I.~{Csabai}, M.~{Doi},
  M.~{Fukugita}, J.~A. {Frieman}, K.~{Glazebrook}, J.~E. {Gunn}, J.~S.
  {Hendry}, G.~{Hennessy}, Z.~{Ivezi{\'c}}, S.~{Kent}, G.~R. {Knapp}, H.~{Lin},
  Y.-S. {Loh}, R.~H. {Lupton}, B.~{Margon}, T.~A. {McKay}, A.~{Meiksin}, J.~A.
  {Munn}, A.~{Pope}, M.~W. {Richmond}, D.~{Schlegel}, D.~P. {Schneider},
  K.~{Shimasaku}, C.~{Stoughton}, M.~A. {Strauss}, M.~{SubbaRao}, A.~S.
  {Szalay}, I.~{Szapudi}, D.~L. {Tucker}, B.~{Yanny}, and D.~G. {York}.
\newblock {Detection of the Baryon Acoustic Peak in the Large-Scale Correlation
  Function of SDSS Luminous Red Galaxies}.
\newblock {\em \apj}, 633:560--574, 2005.

\bibitem{2008PhRvD..77b3533C}
M.~{Crocce} and R.~{Scoccimarro}.
\newblock {Nonlinear evolution of baryon acoustic oscillations}.
\newblock {\em \prd}, 77(2):023533--+, January 2008.

\bibitem{2008PhRvD..77d3525S}
R.~E. {Smith}, R.~{Scoccimarro}, and R.~K. {Sheth}.
\newblock {Motion of the acoustic peak in the correlation function}.
\newblock {\em \prd}, 77(4):043525--+, February 2008.

\bibitem{2008ApJ...686...13S}
H.-J. {Seo}, E.~R. {Siegel}, D.~J. {Eisenstein}, and M.~{White}.
\newblock {Nonlinear Structure Formation and the Acoustic Scale}.
\newblock {\em \apj}, 686:13--24, October 2008.

\bibitem{2008arXiv0812.1392K}
J.~{Kim}, C.~{Park}, J.~R.~I. {Gott}, and J.~{Dubinski}.
\newblock {The Horizon Run N-body Simulation: Baryon Acoustic Oscillations and
  Topology of Large Scale Structure of the Universe}.
\newblock {\em ArXiv e-prints}, December 2008.

\bibitem{1996ApJ...473..620S}
R.~{Scoccimarro} and J.{}A. {Frieman}.
\newblock {Loop Corrections in Nonlinear Cosmological Perturbation Theory. II.
  Two-Point Statistics and Self-Similarity}.
\newblock {\em \apj}, 473:620--644, 1996.

\bibitem{2006PhRvD..73f3520C}
M.~{Crocce} and R.~{Scoccimarro}.
\newblock {Memory of initial conditions in gravitational clustering}.
\newblock {\em \prd}, 73(6):063520--+, March 2006.

\bibitem{2004PhRvL..93q1104K}
J.~{Khoury} and A.~{Weltman}.
\newblock {Chameleon Fields: Awaiting Surprises for Tests of Gravity in Space}.
\newblock {\em Physical Review Letters}, 93(17):171104--+, October 2004.

\bibitem{2004PhRvD..69d4026K}
J.~{Khoury} and A.~{Weltman}.
\newblock {Chameleon cosmology}.
\newblock {\em \prd}, 69(4):044026--+, February 2004.

\bibitem{Mota:2003tc}
David~F. Mota and John~D. Barrow.
\newblock {Varying alpha in a more realistic universe}.
\newblock {\em Phys. Lett.}, B581:141--146, 2004.

\bibitem{2003PhRvD..67f4002L}
A.~{Lue} and G.~{Starkman}.
\newblock {Gravitational leakage into extra dimensions: Probing dark energy
  using local gravity}.
\newblock {\em \prd}, 67:64002, 2003.

\bibitem{2003PhRvD..68b4012D}
G.~{Dvali}, A.~{Gruzinov}, and M.~{Zaldarriaga}.
\newblock {The accelerated universe and the Moon}.
\newblock {\em \prd}, 68(2):024012--+, 2003.

\bibitem{2008PASP..120...20M}
T.~W. {Murphy}, E.~G. {Adelberger}, J.~B.~R. {Battat}, L.~N. {Carey}, C.~D.
  {Hoyle}, P.~{Leblanc}, E.~L. {Michelsen}, K.~{Nordtvedt}, A.~E. {Orin}, J.~D.
  {Strasburg}, C.~W. {Stubbs}, H.~E. {Swanson}, and E.~{Williams}.
\newblock {The Apache Point Observatory Lunar Laser-ranging Operation:
  Instrument Description and First Detections}.
\newblock {\em \pasp}, 120:20--37, January 2008.

\bibitem{2006MNRAS.366..547M}
P.~{McDonald}, H.~{Trac}, and C.~{Contaldi}.
\newblock {Dependence of the non-linear mass power spectrum on the equationof
  state of dark energy}.
\newblock {\em \mnras}, 366:547--556, February 2006.

\bibitem{2008JCAP...10..036P}
M.~{Pietroni}.
\newblock {Flowing with time: a new approach to non-linear cosmological
  perturbations}.
\newblock {\em Journal of Cosmology and Astro-Particle Physics}, 10:36--+,
  October 2008.

\end{thebibliography}

\end{document}